\pdfoutput=1
\documentclass[11pt]{article}
\usepackage{amsmath}
\usepackage{amssymb}
\usepackage{slashed}
\usepackage{graphicx,epstopdf}
\usepackage{cite}
\usepackage{pbox}
\usepackage{geometry}
\usepackage{float}
\usepackage{array}
\usepackage{amsfonts}
\usepackage{graphicx}
\usepackage[font={small}]{caption}
\usepackage[table]{xcolor}
\usepackage{xcolor}
\usepackage{hyperref}
\hypersetup{
    colorlinks=true,
    linkcolor=red,
    filecolor=magenta,      
    urlcolor=cyan,
   citecolor=blue,
} 
\definecolor{RawSienna}{cmyk}{0,0.72,1,0.45}
\definecolor{dgreen}{rgb}{0.0,0.42,0.13}
\definecolor{darkblue}{rgb}{0.0, 0.0, 0.55}
\definecolor{cornellred}{rgb}{0.7, 0.11, 0.11}
\definecolor{calpolypomonagreen}{rgb}{0.08, 0.5, 0.5}

\def\beq{\begin{equation}}
\def\eeq{\end{equation}}
\def\bea{\begin{eqnarray}}
\def\eea{\end{eqnarray}}
\makeatletter
\@addtoreset{equation}{section}

\begin{document}
%\begin{flushright}
%SINP-APC-13/03
%\end{flushright}
%\thispagestyle{empty}
\title{\LARGE \bf Generalized $\mathbb{Z}_2\times \mathbb{Z}_2$ in Scaling neutrino Majorana mass matrix and baryogenesis via  flavored leptogenesis }
\author{{Roopam Sinha\footnote{roopam.sinha@saha.ac.in}, Rome Samanta \footnote{rome.samanta@saha.ac.in}, Ambar Ghosal\footnote{ambar.ghosal@saha.ac.in}
}\\
%a)Department of Physics, Gurudas College,
%Narkeldanga,
% Kolkata-700054, India\\
%a) ???\\
 Saha Institute of Nuclear Physics, HBNI, 1/AF Bidhannagar,
  Kolkata 700064, India \\  
  }
%\keywords{Neutrino Physics, Beyond Standard Model}
\maketitle
\begin{abstract}
We investigate the consequences of a generalized $\mathbb{Z}_2\times\mathbb{Z}_2$ symmetry on a scaling neutrino Majorana mass matrix. It enables us to determine definite analytical relations between the mixing angles $\theta_{12}$ and $\theta_{13}$, maximal CP violation for the Dirac type and vanishing for the Majorana type. Beside the other testable predictions on the low energy neutrino parameters such as $\beta\beta_{0\nu}$ decay matrix element $|M_{ee}|$ and the light neutrino masses $m_{1,2,3}$, the model also has intriguing consequences from the perspective of leptogenesis. With the assumption that the required CP violation for leptogenesis is created  by the decay of  lightest ($N_1$) of the heavy Majorana neutrinos, only $\tau$-flavored leptogenesis scenario is found to be allowed in this model. For a normal (inverted) ordering of light neutrino masses, $\theta_{23}$ is found be less (greater) than its maximal value, for the final baryon asymmetry $Y_B$ to be in the observed range. Besides, an upper and a lower bound on the mass of $N_1$ have also been estimated. Effect of the heavier neutrinos $N_{2,3}$ on final $Y_B$ has been worked out  subsequently. The predictions of this model will be tested in the experiments such as nEXO, LEGEND, GERDA-II, T2K, NO$\nu$A, DUNE etc.
\end{abstract}
\newpage 

\section{Introduction} The neutrino oscillation data, adhering to the bound on the sum of the three electroweak neutrino masses and the results of $\beta\beta_{0\nu}$ decay experiments severely constrain the textures of light neutrino mass matrix. Admissible textures of the mass matrix satisfying the above experimental constraints thus can be tested in future through their predictions regarding the yet unresolved issues such as the hierarchy of neutrino masses, octant determination of $\theta_{23}$, and particularly, CP violation in the leptonic sector which might have implication on the matter-antimatter asymmetry of the universe. Besides, if neutrino is a Majorana  particle, the prediction of Majorana phases will also serve as an added ingredient to discriminate different models. From the symmetry point of view thus it is a challenging task to integrate theoretical considerations involving different symmetry/ansatz in addition to the Standard Model (SM).\\

Recently, the idea of residual symmetry\cite{res,dicus} has attracted much attention to explore the flavor structure of light neutrino mass matrix. In this approach, the neutrino mass matrix is attributed some residual or remnant symmetry of a horizontal flavor group.  It can be shown that the  Majorana type nondegenerate light neutrinos  lead to an invariance of the effective light neutrino  mass matrix  under a $\mathbb{Z}_2\times\mathbb{Z}_2$ symmetry accompanied with a charged lepton mass matrix that enjoys a $\mathbb{Z}_n$ invariance with $n>2$ \cite{res}. Now it is a challenging task to find out larger symmetry groups which  embed these remnant symmetries. Nevertheless, for some predictive residual symmetries, a list of horizontal symmetry  groups has been addressed by Lam\cite{res}. In addtion, viability  of Coxeter groups  as horizontal symmetries in the leptonic sector has been studied recently in Ref.\cite{pal}. Although some of the groups that belong to the Coxeter class have been analyzed in literature (e.g., $S_4$), still there are scopes for a detail study of these groups in the leptonic sector, specifically in the context of grand unified model such as SO(10) that contains Coxeter group as a built-in symmetry\cite{Lam:2014kga}. Furthermore, to constrain the CP violating phases,  a $\mu\tau$-interchange symmetry has been used to implement a nonstandard CP transformation  in Ref.\cite{CP}. Inspired by these well accepted road maps that redirect physicists towards the quest for an ultimate elusive model, in the present work we study the effect of a generalized $\mathbb{Z}_2\times\mathbb{Z}_2$\cite{joshi} that replicates  scaling ansatz\cite{scl,scl2} in conjunction with a  nonstandard CP transformation on light neutrino Majorana mass matrix.\\ 

We first consider a general neutrino mass matrix $M_\nu^0$ with scaling ansatz invariance as an effective low energy symmetry and following residual symmetry approach, interpret the latter as a residual $\mathbb{Z}_2\times\mathbb{Z}_2$ symmetry. Due to the outcome of a vanishing reactor angle $\theta_{13}$ (which is excluded by experiment at more than $5.2\sigma$\cite{th13}), we further use these $\mathbb{Z}_2$ generators to implement CP transformations. Thus instead of an ordinary $\mathbb{Z}_2\times\mathbb{Z}_2$  symmetry,  we now demand a generalized $\mathbb{Z}_2\times\mathbb{Z}_2$ as an effective residual symmetry that extend the scaling ansatz to its complex counterpart. In this case, having a more complicated scaling relationship between its elements, the resultant mass matrices (depending upon the ways of implementation of the symmetry, actually there are two light neutrino mass matrices) are further reconstructed through the type-I seesaw mechanism which incorporates three right chiral singlet neutrino fields $N_{iR}$ $(i=1,2,3)$ in addition to the regular SM field contents.
Although it is nontrivial to combine a flavor and a CP symmetry\cite{Feru1,Holthausen:2012dk,Chen:2014tpa}, a consistent definition for both of them is possible when they satisfy certain condition--usually known as the consistency condition\cite{Holthausen:2012dk,Chen:2014tpa,King:2017guk}. However, at low energy this combined symmetry should be broken to different symmetries in the neutrino and the charged lepton sector, since it is known that at least a common residual CP symmetry in both the sector would imply a vanishing CP violation\cite{Feru1,Holthausen:2012dk,King:2017guk}. Although here we do not focus on the explicit construction of the high energy flavor group, throughout the analysis we assume a diagonal and nondegenerate  charged lepton mass matrix which is protected by a residual symmetry $G_\ell$ after the spontaneous breaking of the combination of CP and flavor symmetry at high energy\cite{Feru1,King:2017guk,Ding:2013hpa}. Depending upon the breaking pattern, there may also be a trivial or a nontrivial CP symmetry in the charged lepton sector\cite{Li:2013jya}. However, as pointed out, the final residual CP symmetries in both the sectors should be different. One can also construct a minimal high energy group from a bottom-up approach knowing the symmetry in the neutrino sector and then finding the symmetry in the charged lepton sector with the automorphism condition as described in Ref \cite{Feru2,Nishi:2016wki}. \\ 

Finally, using the oscillation constraints, tantalizing predictions on the low energy parameters such as neutrino masses, neutrinoless double beta decay, CP violating phases are obtained. Due to the presence of three massive right handed (RH) neutrinos, baryogenesis via leptogenesis scenario is also explored. Interesting conclusions such as octant sensitivity of the atmospheric mixing angle $\theta_{23}$, preconditioned by the observed range of the final baryon asymmetry $Y_B$ and nonoccurrence of unflavored leptogenesis  are also drawn.\\

The paper is organized as follows. Section \ref{s2} contains a brief discussion on residual symmetry and scaling ansatz with a possible modification to the ansatz by extending the former with a nonstandard CP transformation. In section \ref{s3} we discuss a type-I seesaw extension of the analysis made in the previous section. Section \ref{s4} contains a discussion about baryogenesis via leptogenesis scenario related to the present model. In section \ref{s5} we present detail results of the numerical analysis. A discussion on the sensitivity of the heavier neutrinos to the obtained results for the final $Y_B$ is presented in section \ref{s6}. Section \ref{s7} concludes the entire discussion with some promising remarks.

\section{Modification to scaling neutrino mass matrix with generalized $\mathbb{Z}_2\times \mathbb{Z}_2$}\label{s2} 
Before going to an explicit details of our work, let us first discuss the residual $\mathbb{Z}_2\times \mathbb{Z}_2$ symmetry proposed in Ref.\cite{res}. A Majorana neutrino mass matrix $M_\nu$ enjoys a $\mathbb{Z}_2\times \mathbb{Z}_2$ flavor symmetry which can be envisaged as a remnant symmetry of some horizontal flavor group. These horizontal symmetry groups are preferably finite groups since in that case the theory has a more predictive power due to the  discrete number of choices for the residual symmetries $G_i$\cite{res}. A bottom up as well as a top down approach for a viable horizontal group  has been studied in the first one of Ref.\cite{res}. There are plenty of horizontal groups that have been explored in the literature, among them finite groups such as $O_h$\cite{Feru3}, $\mathbb{Z}_m$\cite{Koide:2002cj},  $\mathbb{Z}_m\times \mathbb{Z}_n $\cite{Ghosal:2002mz}, $D_n$\cite{Grimus:2004rj}, $S_4$\cite{Pakvasa:1978tx}, $A_4$\cite{Ma:2001dn}, $\Delta(27)$\cite{Branco:1983tn} and infinite groups such as $SO(3)$ and $SU(3)$\cite{King:2003rf} have drawn much attention.\\  A linear transformation of the neutrino fields $\nu_{L\alpha}\to G_{\alpha\beta}\nu_{L\beta}$ leads to an invariance of an effective neutrino Majorana mass term
\bea
-\mathcal{L}^\nu_{mass}=\frac{1}{2}\bar{\nu}^C_{L\alpha}(M_\nu)_{\alpha \beta}\nu_{L \beta}+{\rm h.c.},
\eea
if the mass matrix $M_\nu$ satisfies the invariance equation
\bea
G^TM_\nu G=M_\nu.\label{1}
\eea
Here $G$ is a $3\times 3$ unitary matrix in flavor basis. It has been shown in Ref.\cite{res} that if an unitary matrix $U$  diagonalizes $M_\nu$ then the  matrix $U^\prime=GU$ also does so  where $U^\prime$ satisfies the condition 
\bea
GU=Ud~~{\rm with}~~d_{lm}=\pm \delta_{lm}.\label{z2d}
\eea 
Among the eight possible choices for $d$, only two of them can be shown to be independent on account of the relation $d_a d_b=d_bd_a=d_c$, which implies $G_a G_b=G_b G_a=G_c$ with $a\neq b\neq c$. These two independent $G$ matrices define a $\mathbb{Z}_2\times\mathbb{Z}_2$ symmetry since $d^2=G^2=I$ as dictated by Eq.\eqref{z2d}. Thus given a mass matrix $M_\nu$, one can obtain $U$ consistent with the symmetries of $M_\nu$. From which $G_a$'s can be obtained as 
 \begin{equation}
G_{a}=U d_{a}U^\dag
\end{equation} 
with $a=1,2,3$. Since $G^2=I$ implies ${\rm det}~G=\pm 1$,  one can choose the independent $d$ matrices corresponding to any value for the determinant of the $G$ matrices. Here without loss of generality, we choose to proceed with ${\rm det}~G=+ 1$ that corresponds to the structure of $d$ matrices as $d_1={\rm diag}~(1,-1,-1)$, $d_2={\rm diag}~(-1,1,-1)$ and $d_3=d_1d_2$. 

Basically for an arbitrary mixing matrix $U$, one can construct a unique $G$, however the reverse is not true due to the degeneracies in the eigenvalues of $d_a$ matrices. From this point, the implementation of the residual symmetry to the neutrino mass matrix takes different paths. Given a leading order mixing matrix, e.g. $U^{\mu\tau}$, construction of $G$ matrices are unique, then for a particular $G$ matrix, one might or might not have $U^{\mu\tau}$. Papers such as \cite{dicus} discuss scenarios like soft breaking of one of the two residual symmetries such that presence of the other with its degenerate eigenvalues enhances the degrees of choice of the mixing matrix in accordance with the phenomenological requirement. On the other hand, in Ref.\cite{CP,joshi,val}, as a more predictive scenario, invariance of the neutrino mass matrix under an extended $\mu\tau$ symmetry (${\rm CP}^{\mu\tau}$ or CP transformation with the $\mu\tau$-symmetry) has been considered. Both the schemes have their own uniqueness in terms of the predictions on the low energy neutrino parameters. However, in this work, we follow the second approach due to its robust predictions on CP violating phases which are related to the matter antimatter asymmetry of the universe\cite{planck}.\\

We interpret the Strong Scaling Ansatz (SSA) proposed in Ref.\cite{scl}, as a residual $\mathbb{Z}_2\times \mathbb{Z}_2$ symmetry. Since SSA leads to a vanishing $\theta_{13}$, a possible modification to this has been made by generalizing the two independent ordinary $\mathbb{Z}_2$ invariance to their complex counterpart, i.e., two independent $\mathbb{Z}_2^{\rm CP}$ invariance. Thus the SSA has been extended to its complex version by means of a generalized $\mathbb{Z}_2\times \mathbb{Z}_2$ symmetry (see Ref.\cite{joshi} for another such extension in case of TBM mixing). Let's discuss now the exact methodology of our analysis:\\

We consider a column wise scaling relations in the elements of $M^{0}_\nu$ in flavor space as
\begin{equation}
\frac{(M_\nu^{0})_{e\mu}}{(-M_\nu^{0})_{e\tau}}=\frac{(M_\nu^{0})_{\mu\mu}}{(-M_\nu^{0})_{\mu\tau}}=\frac{(M_\nu^{0})_{\tau\mu}}{(-M_\nu^{0})_{\tau\tau}}=k,
\label{scr}
\end{equation}
 where $k$ is a real and positive dimensionless scaling factor. The superscript `0' on $M_\nu$ symbolizes SSA as a leading order matrix in this analysis. Now the structure for $M_\nu^{0}$ dictated by the ansatz of  Eq.\eqref{scr} comes out as
 \begin{equation}
M_\nu^{0}=\begin{pmatrix}
P & -Qk & Q\\-Qk & Rk^2 & -Rk\\Q & -Rk & R
\end{pmatrix}.\label{mnusrs}
\end{equation} 
Here $P,Q, R$ are a priori unknown, complex mass dimensional quantities. The minus sign in Eq.\eqref{scr} has been considered to be in conformity with the PDG convention\cite{pdg}. The matrix in Eq.\eqref{mnusrs} is diagonalized by a unitary matrix $U^0$  having a form
\bea
U^{0}=\begin{pmatrix}
c_{12}^0&s_{12}^0e^{i\alpha}&0\\
-\frac{ks_{12}^0}{\sqrt{1+k^2}}&\frac{kc_{12}^0}{\sqrt{1+k^2}}e^{i\alpha /2}&\frac{1}{\sqrt{1+k^2}}e^{i\beta /2}\\
\frac{s_{12}^0}{\sqrt{1+k^2}}&-\frac{c_{12}^0}{\sqrt{1+k^2}}e^{i\alpha /2}&\frac{k}{\sqrt{1+k^2}}e^{i\beta /2}\\

\end{pmatrix},\label{u0}
\eea
where  $c_{12}^0=\cos\theta_{12}^0$, $s_{12}^0=\sin\theta_{12}^0$ which are calculated in terms of the parameters of $M_\nu^{0}$, and $\alpha,\beta$ represents the Majorana phases.  
SSA predicts a vanishing $\theta_{13}$ (hence no measurable leptonic Dirac CP-violation) as one can see from Eq.\eqref{u0} and an inverted neutrino mass ordering (i.e., $m_{2,1}>m_3$), with $m_3=0$. As previously mentioned,  one needs to modify the ansatz to generate a non-zero $\theta_{13}$. Now using the  paradigm of residual symmetry  as described in the earlier part of this section, one can calculate the $G_a$ matrices using the relation 
\bea
G_a^{(k)}=U^{0}d_aU^{0\dagger}\label{srsg}
\eea
with $G_a^{(k)}$ as the $\mathbb{Z}_2$ generators for a scaling ansatz invariant $M_\nu$. Similar to Eq.\eqref{1}, $M_\nu^{0}$ will then satisfy the invariance equation
\bea
\Big(G_a^{(k)}\Big)^T M^0_\nu G_a^{(k)}=M^0_\nu.\label{srsinvr}
\eea
Now using Eq.\eqref{srsg} we calculate the corresponding $G_{a}^{(k)}$ ($a=1,2,3$) matrices and present them as
\begin{equation}G_1^{(k)}=\begin{pmatrix}
\cos 2\theta^0_{12} & -k(1+k^2)^{-1/2}\sin 2\theta^0_{12} & -(1+k^2)^{-1/2}\sin 2\theta^0_{12}\\-k(1+k^2)^{-1/2}\sin 2\theta^0_{12} & -(1+k^2)^{-1}(k^2\cos2\theta^0_{12}+1)
& -k(1+k^2)^{-1}(1-\cos 2\theta^0_{12})\\-(1+k^2)^{-1/2}\sin 2\theta^0_{12} & -k(1+k^2)^{-1}(1-\cos 2\theta^0_{12}) & -(1+k^2)^{-1}(k^2+\cos 2\theta^0_{12}) \end{pmatrix},\end{equation}

\begin{equation}G_2^{(k)}=\begin{pmatrix}
-\cos 2\theta^0_{12} & k(1+k^2)^{-1/2}\sin 2\theta^0_{12} & -(1+k^2)^{-1/2}\sin 2\theta^0_{12}\\k(1+k^2)^{-1/2}\sin 2\theta^0_{12} & (1+k^2)^{-1}(k^2\cos2\theta^0_{12}-1)
& -k(1+k^2)^{-1}(1+\cos 2\theta^0_{12})\\-(1+k^2)^{-1/2}\sin 2\theta^0_{12} & -k(1+k^2)^{-1}(1+\cos 2\theta^0_{12}) & -(1+k^2)^{-1}(k^2-\cos 2\theta^0_{12}) \end{pmatrix},\end{equation}  

\begin{equation}G^{(k)}_3=\begin{pmatrix}
-1 & 0 & 0\\0 & (1-k^2)(1+k^2)^{-1}
& 2k(1+k^2)^{-1}\\0 & 2k(1+k^2)^{-1} & -(1-k^2)(1+k^2)^{-1} \end{pmatrix}.\label{g3ks}
\end{equation}
Note that all the $G^{(k)}_a$ matrices are symmetric by construction. Now to modify SSA, we generalize this $\mathbb{Z}_2\times\mathbb{Z}_2$ by  implementing  CP transformations on the neutrino fields\cite{CPt} with the $\mathbb{Z}_2$ generators ($G_a^{(k)}={G_a^{(k)}}^T$) as\footnote{The matrices that represent the CP symmetry should be symmetric\cite{Feru1}.}
\bea
\nu_{L\alpha}\to i(G_a^{(k)})_{\alpha\beta}\gamma^0\nu^C_{L\beta}.
\eea
 This extends the real horizontal invariance of $M_\nu^0$ in Eq.\eqref{srsinvr} to its complex counterpart, i.e.
\bea
\Big(G_a^{(k)}\Big)^TM_\nu G_a^{(k)}=M_\nu^*.
\eea
Therefore the SSA, elucidated as a $\mathbb{Z}_2\times\mathbb{Z}_2$ symmetry, has now been modified to an extended SSA, interpreted as a   complex $\mathbb{Z}_2\times\mathbb{Z}_2$ symmetry which is some time also referred  as a generalized $\mathbb{Z}_2\times\mathbb{Z}_2$ symmetry of $M_\nu$\cite{joshi}. In the next subsections we show that there are only two ways in which such a complex extension can be done. 

%,\hspace{0.5cm}N_{Ri}\to i(G_R)_{ij}\gamma^0N^C_{Rj}
%and demanding the invariance relations \begin{equation}
%G_R^\dag m_D G_L=m_D^*,\hspace{0.5cm} G_R^\dag M_R G_R^*=M_R^*\label{com}
%\end{equation} Eqn \ref{} and \ref{} jointly imply \begin{equation}
%G_L^TM_\nu G_L=M_\nu^*\end{equation}

\subsection{Case I: Complex extension of $G_{2,3}^{(k)}$ Invariance}
The complex invariance relations of $M_\nu$ related to $G_{2,3}^{(k)}$ is now written as
\begin{equation}
\Big(G_{2,3}^{(k)}\Big)^TM_\nu G_{2,3}^{(k)}=M_\nu^*,\label{g23c}
\end{equation}which in turn implies 
\begin{equation}
\Big(G_{1}^{(k)}\Big)^T M_\nu G_{1}^{(k)}=M_\nu\label{g1r}
\end{equation} 
owing to the closure property of the $G^{(k)}_a$ ($a=1,2,3$)  matrices.\\

Eq.\eqref{g23c} leads to a most general Majorana neutrino mass matrix of the form \begin{equation}
M_{\nu}^{MS1}=\begin{pmatrix}
p & -q_1k+i\frac{q_2}{k}  & q_1+iq_2\\-q_1k+i\frac{q_2}{k} & r-\frac{s(k^2-1)}{k}+i\frac{2q_2\kappa_+}{\sqrt{1+k^2}} &s+i\frac{q_2\kappa_+(k^2-1)}{k\sqrt{1+k^2}}\\q_1+iq_2 & s+i\frac{q_2\kappa_+(k^2-1)}{k\sqrt{1+k^2}} & r-i\frac{2q_2\kappa_+}{\sqrt{1+k^2}}
\end{pmatrix}\label{g23p}
\end{equation}
 with 
 \begin{equation}
r=(sk+p)-q_1\sqrt{1+k^2}(\kappa_+-\frac{1}{\kappa_+}),
\end{equation}
\bea
\kappa_+=(\cot 2\theta_{12}^0+{\rm cosec}2\theta_{12}^0).
\eea
%\begin{equation}
%z_2=-\frac{2y_2}{\sqrt{1+k^2}}(\cot 2\theta^0_{12}+ \text{cosec}\hspace{0.2cm}2\theta^0_{12}).
%\end{equation} 
Here $p$, $q_{1,2}$, $r$ and $s$ are real, mass dimentional quantities and the superscript `$MS$' stands for `Modified Scaling'. It has  already been shown in Ref.\cite{Rome} that $(G_{3}^{(k)})^TM_\nu G_{3}^{(k)}=M_\nu^*$ leads to the results \begin{equation}
\tan \theta_{23}=k^{-1}, \label{th23k}
\end{equation}\begin{equation}
\sin\alpha=\sin\beta=\cos\delta=0.
\end{equation} 
Now in the present case,  the overall real $G_1^{(k)}$ (cf. Eq.\eqref{g1r}) invariance of $M_\nu$ fixes the first column of $U_{PMNS}$ to the first column of $U^{0}$. Therefore, one gets the relation between the solar and the reactor mixing angle as
\begin{equation}
|\cos\theta_{12}\cos\theta_{13}|=\cos\theta^0_{12}\Rightarrow\sin^2\theta_{12}=1-\cos^2\theta^0_{12}(1+\tan^2\theta_{13}). \label{th20}
\end{equation}
\subsection{Case II: Complex extension of $G_{1,3}^{(k)}$ Invariance}
In this case, the complex invariance relations of $M_\nu$ due to $G_{1,3}^{(k)}$ can be written as 
\begin{equation}
\Big(G_{1,3}^{(k)}\Big)^TM_\nu G_{1,3}^{(k)}=M_\nu^*,\label{g13c}
\end{equation} 
which leads to 
\begin{equation}
\Big(G_{2}^{(k)}\Big)^T M_\nu G_{2}^{(k)}=M_\nu.\label{g2r}
\end{equation} 
Eq.\eqref{g13c} leads to the mass matrix $M_\nu^{MS2}$ having a form same as  $M_\nu^{MS1}$ as given in Eq.\eqref{g23p} where $\kappa_+$ is replaced with $\kappa_-=-1/\kappa_+$. Similar to the previous case, a complex invariance due to $G_{3}^{(k)}$ leads to the predictions
\begin{equation}
\tan \theta_{23}=k^{-1},\label{th23k2}
\end{equation}
\begin{equation}
\sin\alpha=\sin\beta=\cos\delta=0.
\end{equation} 
Now the overall real $G_2^{(k)}$ (cf. Eq.\eqref{g2r}) invariance of $M_\nu$ fixes the second column of $U_{PMNS}$ to the second column of $U^{0}$ which gives rise to a relation between the solar and the reactor mixing angle as
\begin{equation}
|\sin\theta_{12}\cos\theta_{13}|=\sin\theta^0_{12}\Rightarrow\sin^2\theta_{12}=\sin^2\theta^0_{12}(1+\tan^2\theta_{13}). \label{th1201}
\end{equation}
%\subsection{Case III: Complex extension of $G_{1,2}^{(k)}$ Invariance}
%Similar to the previous cases, the complex invariance relations of $M_\nu$ related to $G_{1,2}^{(k)}$ can be written as
%\begin{equation}
%(G_{1,2}^{(k)})^TM_\nu G_{1,2}^{(k)}=M_\nu^*,\label{g12c}
%\end{equation}
%which in turn implies \begin{equation}
%(G_{3}^{(k)})^T M_\nu G_{3}^{(k)}=M_\nu.\label{g3 r}
%\end{equation}  
%However, since $G_{3}^{(k)}$ fixes the third column of $U_{PMNS}$ to the third column of $U^{0}$, this invariance leads to a vanishing value of $\theta_{13}$. Thus simultaneous complex invariance of $M_\nu$ due to $G_{1}^{(k)}$ and $G_{2}^{(k)}$, is not a viable extension of SSA.\\
Similar to the previous cases, complex invariance due to $G_{1,2}^{(k)}$ leads to an overall real invariance due to $G_{3}^{(k)}$ which leads to a vanishing $\theta_{13}$. Thus this is a case of least interest. For both the viable cases, we determine three CP phases ($\cos\delta=0,\alpha,\beta=0$ or $\pi$). Thus there are 6 real free parameters  $p, q_{1,2},s ,k$ and $\kappa_+$ (or $\theta^0_{12}$) (cf. Eq.\eqref{g23p}) in both the mass matrices. However, one can trivially track the parameters $k$ and $\theta_{12}^0$ on account of the relations in \eqref{th23k} or \eqref{th23k2} and \eqref{th20} or \eqref{th1201}.  Thus the other four parameters account for one mixing angle and three neutrino masses. However, to fix the absolute neutrino mass scale,  we additionally use some constraints from baryogenesis  as discussed in the numerical section.\\

We note that the prediction of the CP phases in the extended SSA scheme are identical to the case of ${\rm CP}^{\mu\tau}$\cite{CP}. Therefore the question arises how one might distinguish the ${\rm CP}^{\mu\tau}$ and the extended SSA experimentally? First of all, both the Strong Scaling Ansatz (SSA) and the $\mu\tau$ symmetry lead to $\theta_{13}=0$ at the leading order and therefore, has to be abandoned. However, one can in principle differentiate SSA from the $\mu\tau$ reflection symmetry via their predictions of atmospheric mixing angle $\theta_{23}$. The former in general predicts a nonmaximal $\theta_{23}$ (for $k\neq 1$) given by $\theta_{23}=\tan^{-1}(k^{-1})$  while a maximal value ($\theta_{23}=\pi/4$) is predicted by the latter.\\

Furthermore, in the extended scheme, besides the similar predictions for the CP phases an arbitrary nonvanishing value of the reactor mixing angle $\theta_{13}$ is predicted in both the cases (extended SSA and ${\rm CP}^{\mu\tau}$). However, the prediction on the $\theta_{23}$ is different for each case. Interestingly, even after the extension, the value of $\theta_{23}$ survives for both the cases i.e., $\theta_{23}=\tan^{-1}(k^{-1})$ for the  SSA as well as extended SSA and $\theta_{23}=\pi/4$ for $\mu\tau$ symmetry and its extended version (${\rm CP}^{\mu\tau}$). If experiments find a nonmaximal $\theta_{23}$ at a significant confidence level (recently there is a hint from NO$\nu$A regarding the nonmaximality of $\theta_{23}$ at 2.6 $\sigma$ CL\cite{Adamson:2017qqn}) then the ${\rm CP}^{\mu\tau}$ symmetry will be ruled out while our proposal of an extended SSA (that predicts a nonmaximal $\theta_{23}$ in general) will continue to survive.\\

Before proceeding further we should comment on the fulfillment of the consistency conditions\cite{Holthausen:2012dk,Chen:2014tpa,King:2017guk} as mentioned in the introduction. Here we have discussed two cases. In the first one $G_{2,3}^{(k)}$ are the CP symmetries which further result in a $G_{1}^{(k)}$ invariance of the mass term while in the second case, the CP generators $G_{1,3}^{(k)}$ lead to an invariance of the mass term due to the $G_{2}^{(k)}$. Now the consistency condition in case of a $\mathbb{Z}_2$ group implies\cite{King:2017guk} 
\bea
X_r\rho_r^*(g)X_r^{-1}=\rho_r(g),\label{const}
\eea
where  $X_r$ is a unitary matrix representing CP symmetry  which acts on a generic multiplet $\varphi$  as
\bea
X_r\varphi(x)\xrightarrow{\text{CP}} X_r\varphi(x^\prime)
\eea
with $x^\prime=(t,-{\bf x})$ and $\rho_r(g)$ is a representation for the element $g$ of the flavor group in an irreducible representation ${\bf r}$. In our analysis, $G_i^{(k)}$'s are real, and hence, the condition in Eq.\eqref{const} turns out to be 
\bea
G_{2,3}^{(k)} G_1^{(k)} (G_{2,3}^{(k)} )^{-1}=G_1^{(k)}~~~{\rm for}~~{\rm Case~I};\nonumber \\
G_{1,3}^{(k)} G_2^{(k)} (G_{1,3}^{(k)} )^{-1}=G_2^{(k)}~~~{\rm for}~~{\rm Case~II}.
\eea 
Since $\Big({G_i^{(k)}}\Big)^2=1$, $\Big({G_i^{(k)}}\Big)^{-1}=G_i^{(k)}$ and each $G_i^{(k)}$ commutes with each other, the consistency condition is trivially satisfied for both the cases. However the main challenge is to ensure that such conditions are fulfilled for the larger (embedding) symmetries\cite{Feru1,Holthausen:2012dk,Chen:2014tpa} which we do not explore here in this work.\\

Resolving the shortcomings of SSA, both the viable modified SSA matrices, referred as $M_\nu^{MS1}$ and $M_\nu^{MS2}$,  possess intriguing phenomenology. This has been discussed in section \ref{s5} on numerical analysis. For the time being let's focus on the implementation of the  symmetry in a more specific way. So far we have discussed a possible complex extension for a general $M_\nu$, not so about the origin of the neutrino masses. This would be interesting to see the effects of generalized $\mathbb{Z}_2\times\mathbb{Z}_2$ on a particular mechanism that generates the light neutrino masses. Obviously, the choice  depends upon the phenomenological interest. Here we choose  the type-I seesaw mechanism and investigate possible consequences of the generalized $\mathbb{Z}_2\times\mathbb{Z}_2$ to explore the phenomena of baryogenesis via leptogenesis. A detailed discussion about these has been presented in the next two sections. First, we show the reconstruction of the effective modified SSA matrices through type-I seesaw mechanism with  proper implementation of the symmetry on the constituent matrices ($m_D$ and $M_R$). Then we discuss  some aspects of baryogenesis via leptogenesis related to this scheme. 
\section{Reconstruction of modified scaling matrices with type-I seesaw} \label{s3}
For the realization of generalized $\mathbb{Z}_2\times\mathbb{Z}_2$ in the context of type-I seesaw mechanism, we define two separate `$G$' matrices $G_L$ and $G_R$ for $\nu_L$ and $N_R$ fields respectively. Now  the CP transformations are  defined on these fields as\cite{chen2}  \bea
\nu_{L\alpha }\rightarrow i (G_L)_{\alpha \beta}\gamma^0 \nu_{L\beta}^C, \hspace{.5cm}
N_{R\alpha }\rightarrow i (G_R)_{\alpha \beta}\gamma^0 N_{R\beta}^C. \label{seCP}
\eea
With $m_D$ as a Dirac type and $M_R$ as a diagonal nondegenerate Majorana type mass matrix, the Lagrangian for type-I seesaw 
\bea
-\mathcal{L}= \bar{N}_{iR} (m_D)_{i\alpha}l_{L\alpha}+\frac{1}{2}\bar{N}_{iR}(M_R)_i \delta _{ij}N_{jR}^C + {\rm h.c}. \label{selag}
\eea
leads to the effective $3\times3$ light neutrino Majorana mass matrix $M_\nu$ as
\bea
M_\nu=-m_D^TM_R^{-1}m_D. \label{swmnu}
\eea
Now the invariance of the mass terms of Eq.\eqref{selag} under the CP transformations defined in Eq.\eqref{seCP} leads to the relations
\bea
G_R^{\dagger}m_D G_L=m_D^*, \hspace{.3cm} G_R^{\dagger}M_R G_R^*=M_R^*. \label{trmdmr}
\eea
Eqs.\eqref{swmnu} and \eqref{trmdmr} together imply $G_L^TM_\nu G_L=M_\nu^*$. Now, specifying $G_L$ by $G_i^{(k)}$, we  obtain the key equation 
\bea
\Big(G_i^{(k)}\Big)^T M_\nu G_i^{(k)}=M_\nu^*. 
\eea 
Since $M_R$ is taken to be diagonal i.e., $M_R={\rm diag\hspace{.5mm}} (M_1,M_2,M_3)$, the corresponding symmetry generator matrix $G_R$ is diagonal\cite{chen2} with entries $\pm 1$, i.e., 
\begin{equation} 
(G_R)_{lm}=\pm\delta_{lm}.\label{gr1d}
\end{equation}
 which implies for each $G_L$, there are eight different structures for $G_R$ that correspond to eight different choices of $m_D$. However, a straightforward computation shows that for the case-I, the $G_R$ matrix compatible with $G_2^{(k)}$ and $G_3^{(k)}$ should be taken as $(G_{R})_2=\text{diag}\hspace{1mm}(1,1,1)$ and $(G_{R})_3=\text{diag}\hspace{1mm}(-1,- 1,- 1)$ respectively. Similarly for Case-II also, those are taken as $(G_{R})_1=\text{diag}\hspace{1mm}(1,1,1)$ and  $(G_{R})_3=\text{diag}\hspace{1mm}(-1,- 1,- 1)$ for $G_1^{(k)}$ and $G_3^{(k)}$. It can be shown that all the other choices of $G_R$ are incompatible with scaling symmetry. Therefore, the first of Eq.\eqref{trmdmr} leads to
\bea
m_D G_3=-m_D^*,m_D G_{2}=m_D^*\label{cons}\hspace{1cm}~{\rm for}~\text{Case-I}\nonumber\\
m_D G_3=-m_D^*,m_D G_{1}=m_D^*\label{cons}\hspace{1cm}~{\rm for}~\text{ Case-II}.
\eea
For both the cases as discussed above, the most general form of $m_D$ that satisfies the constraints of Eq.\eqref{cons} can be parameterized as
\begin{equation}
m_D^{MS}=\begin{pmatrix}
a & b_1+ib_2 & -b_1/k+ib_2k\\e & c_1+ic_2 & -c_1/k+ic_2k\\f & d_1+id_2 & -d_1/k+id_2k
\end{pmatrix}\label{mdcs}
\end{equation} 
with
\begin{equation}
b_1=\pm ak({1+k^2})^{-1/2}\kappa_{\pm},  \label{ss1}
\end{equation}
 \begin{equation} c_1=\pm ek({1+k^2})^{-1/2}\kappa_{\pm}, \label{ss2}
  \end{equation}
 \begin{equation} 
 d_1=\pm fk({1+k^2})^{-1/2}\kappa_{\pm} . \label{ss3}
 \end{equation}
 Here the  `$\pm$' sign in the expressions of $b_1,c_1$ and $d_1$ are for Case-I and Case-II respectively.
In Eq.\eqref{mdcs} $a,e,f,b_{2},c_{2}$ and $d_{2}$ are six a priori unknown real mass dimensional quantities and $k$ is a real, positive, dimensionless parameter. Now using the seesaw relation in Eq.\eqref{swmnu}, it is easy to reconstruct the effective mass matrices $M_\nu^{MS1}$ and $M_\nu^{MS2}$ for Case-I and Case-II respectively. In Table \ref{mdp}, we present the parameters of the  effective light neutrino mass matrix in terms of the Dirac and Majorana components.
\begin{table}[H]
\caption{Parameters of $M_\nu$.}  \label{mdp}
%\centering 
\begin{center}
\begin{tabular}{|c|}
\hline
$ p=-(\frac{a^2}{M_1}+\frac{e^2}{M_2}+\frac{f^2}{M_3})$\\
%\hline
$q_1=-\frac{\kappa_\pm p}{\sqrt{1+k^2}}$\\
%\hline
$q_2=-k(\frac{ab_2}{M_1}+\frac{ec_2}{M_2}+\frac{fd_2}{M_3})$\\
%\hline
$s=-\frac{\kappa_\pm^2 p k}{1+k^2}+k(\frac{b_2^2}{M_1}+\frac{c_2^2}{M_2}+\frac{d_2^2}{M_3})$\\
%\hline
$r=(sk+p)-q_1\sqrt{1+k^2}(\kappa_\pm-\frac{1}{\kappa_\pm})$\\
%\hline
%$z_2=-\frac{2\kappa y_2}{\sqrt{1+k^2}}$\\
\hline
\end{tabular}
\end{center}
%\label{table:fundam} % is used to refer this table in the text
\end{table}
\noindent 
Once again `$\pm$' sign in $\kappa$ are for Case-I and Case-II respectively. \\

Before concluding this section we would like to address the following: It clear from  Eq.\eqref{g3ks} and Eq.\eqref{gr1d}) that the matrices $G_L$ and $G_R$ are of different form. This is since we choose to work in a basis where $M_R$ is diagonal but $m_D$ is not (``leptogenesis basis"\cite{chen2}). However that does not mean that the left handed and right handed field must transform  differently. The form of $G_R$, i.e., $G_R={\rm diag}~(\pm 1,\pm 1,\pm 1)$ is obtained purely for the diagonal $M_R$ matrix. In principle one may assume same residual symmetry (say $G$) in the matrices $m_D$ and $M_R$ when both of them are nondiagonal. However, in a basis where $M_R$ is diagonal the symmetry in the nondiagonal $M_R$  ultimately changes to $G_R={\rm diag}~(\pm 1,\pm 1,\pm 1)$ while the symmetry in the left handed field remains the same. \\

To see this explicitly, we consider the Lagrangian of Eq.\eqref{selag} with a nondiagonal $M_R$. Now $M_R$ could be diagonalized by a unitary matrix $U_N$ as
\bea
U_N^\dagger M_R U_N=M_R^d={\rm diag} ~ (M_1, M_2, M_3), \label{mrdiag}
\eea
where $M_R^d$ is a real diagonal matrix with nondegenerate eigenvalues. Eq.\eqref{trmdmr} can now be rewritten  as
\bea
G^{\dagger}m_D G=m_D^*, \hspace{.3cm} G^{\dagger}M_R G^*=M_R^*, \label{trmdmr1}
\eea
where we have assumed same symmetry for both the fields.
Now the second equation of Eq.\eqref{trmdmr1} and Eq.\eqref{mrdiag} together imply
\bea
U_N^T G^\dagger U_N= d^\dagger, \label{rfc1}
\eea
where $d$ is a diagonal matrix with $d_{jj}=\pm 1$. In the basis where the RH neutrino mass matrix is diagonal one can have a modified Dirac matrix as 
\bea
m_D\rightarrow m_D^\prime = U_N^\dagger m_D.\label{rfc2}
\eea
 Thus the first equation of Eq.\eqref{trmdmr1} and Eq.\eqref{rfc1} give 
\bea
U_N^* d^\dagger U_N^\dagger m_D G =m_D^* ~~{\rm or}~~d^\dagger m_D^\prime G = m_D^{\prime *},
\eea
where $m_D^\prime$ is defined in Eq.\eqref{rfc2}. Thus starting from a basis where $M_R$ is nondiagonal, we obtain the identical  complex symmetry condition on the Dirac mass matrix as given in Eq.\eqref{trmdmr} in the basis where $M_R$ is diagonal. This is worth mentioning that the matrix $d$ is basically the matrix $G_R$ of Eq.\eqref{gr1d} since they both are diagonal with entries $\pm1$. 

\section{Baryogenesis via leptogenesis}\label{s4}
Baryogenesis via leptogenesis\cite{fuku,fuku2} is a phenomena where CP violating and out of equilibrium decays from heavy Majorana neutrinos generate a lepton asymmetry which is thereafter converted into baryon asymmetry by sphaleron transition\cite{hooft}. The pertinent Lagrangian for the process can be written as
\bea
 -\mathcal{L}=\lambda_{i\alpha} \bar{N}_{Ri}\tilde{\phi}^\dag l_{L\alpha}+\frac{1}{2}\bar{N}_{iR}(M_R)_i \delta _{ij}N_{jR}^C+{\rm h.c.}\label{cpa}
 \eea
 where $l_{L\alpha}=\begin{pmatrix}\nu_{L\alpha} & e_{L\alpha}\end{pmatrix}^T$ is the SM lepton doublet of flavor $\alpha$, and $\tilde{\phi}=i\tau_2\phi^*$ with $\phi=\begin{pmatrix}\phi^+ & \phi^0\end{pmatrix}^T$ being the Higgs doublet. Thus the possible decays of $N_i$ from Eq.\eqref{cpa} are $N_i\to e^-_\alpha\phi^+$, $N_i\to \nu_\alpha\phi^0$, $N_i\to e_\alpha^+\phi^-$, and $N_i\to \nu^C_\alpha \phi^{0*}$. %The flavour dependent CP-asymmetry paameter is given by the standard formula\begin{equation} \epsilon_i^\alpha=\frac{\Gamma(N_i\to L_\alpha\phi)-\Gamma(N_i\to L_\alpha^C\phi^\dag)}{\Gamma(N_i\to L_\alpha\phi)-\Gamma(N_i\to L_\alpha^C\phi^\dag)},\label{cpa1} \end{equation} where $\Gamma$ represents the partial decay widths. A nonzero value of $\epsilon_i^\alpha$ needs to arise out of the interference between the tree level and one loop contributions[]. This is since at the tree level we have \begin{equation} \Gamma^{tree}(N_i\to L_\alpha\phi)=\Gamma^{tree}(N_i\to L_\alpha^C\phi^\dag)=\frac{M_i}{16\pi}\lambda^{\dag}_{i\alpha}\lambda_{i\alpha}, \text{no sum over i}. \end{equation} One loop contributions come both from vertex correction and self-energy terms. 
The CP asymmetry parameter $\varepsilon_i^\alpha$  that accounts for the required CP violation, arises due to the interference between the tree level, one loop self energy, one loop vertex $N_i$-decay diagrams\cite{fuku} and has a  general expression\cite{pila}
\begin{equation}
\varepsilon_i^\alpha=\frac{1}{4\pi v^2 h_{ii}}\sum\limits_{j\neq i}\left\{\text{I}m[h_{ij}(m_D)_{i\alpha}(m_D^*)_{j\alpha}]g(x_{ij})+\frac{\text{Im}[h_{ji}(m_D)_{i\alpha}(m_D^*)_{j\alpha}]}{1-x_{ij}}\right\}\label{asymp}
\end{equation} 
where $h\equiv m_D m_D^\dag$, $\langle\phi^0\rangle=v/\sqrt{2}$ so that $m_D = v\lambda/\sqrt{2}$, and $x_{ij}=M_j^2/M_i^2$. Furthermore, the loop function
$g(x_{ij})$ has the expression
\begin{equation}
g(x_{ij})=\frac{\sqrt{x_{ij}}}{1-x_{ij}}+f(x_{ij})\label{cpa2}
\end{equation} 
with
\begin{equation}
f(x_{ij})=\sqrt{x_{ij}}\Big[1-(1+x_{ij})\ln\Big(\frac{1+x_{ij}}{x_{ij}}\Big) \Big].
\end{equation} 

%Let us explore some of the physics contained in Eqn.\ref{cpa1}. Depending upon the temperature regime in which leptogenesis occurs, lepton flavors may be fully distinguishable, partly distinguishable or indistinguishable[]. It is reasonable to assume that leptogenesis takes place at $T\sim M_1$. It is known [] that lepton flavors cannot be treated separately if the concerned process occurs above a temperature $T\sim M_1>10^{12}$ GeV. In case the said temperature is lower, two possibilities arise. When $T\sim M_1<10^9$ GeV, all three flavors $(e,\mu,\tau)$ are individually active and we need three CP asymmetry parameters $\epsilon^{e}_i,\epsilon^{\mu}_i$ and $\epsilon^{\tau}_i$ for each generation of RH neutrinos. On the other hand when we have $10^9$ GeV $<T\sim M_1 <10^{12}$ GeV, only the $\tau-$flavor can be identified separately while the $e$ and $\mu$ act indistinguishably. Here we need two CP asymmetry parameters $\epsilon_i^{(2)}=\epsilon^{e}_i+\epsilon^{\mu}_i$, and $\epsilon_i^\tau$ for each of the RH neutrinos. 
Before going to the explicit calculation of $\varepsilon_i^\alpha$ related to this model, let's address some important issues related to leptogenesis. For a hierarchical scenario, e.g., $M_3\gg M_2\gg M_1$, it can be shown that only the decays of $N_1$ matter for the creation of lepton asymmetry while the latter created from the heavier neutrinos get washed out\cite{bari}. Obviously there are certain circumstances when the decays of $N_{2,3}$ are also significant\cite{n2lp}. Again, flavor plays an important role in the phenomena of leptogenesis\cite{abada}. Assuming the temperature scale of the process $T\sim M_1$,  the rates of the Yukawa interaction categorize leptogenesis  into three categories. 1) $T\sim M_1>10^{12}$ GeV, when all interactions with all flavors are out of equilibrium: unflavored leptogenesis. In this case all the flavors are indistinguishable and thus the total CP asymmetry is a sum over all flavors, i.e., $\varepsilon_i=\sum_{\alpha}\varepsilon^\alpha_i$. 2)  $10^9$ GeV $<T\sim M_1 <10^{12}$ GeV, when only the $\tau$ flavor is in equilibrium: $\tau$-flavored leptogenesis. In this regime there are two relevant CP asymmetry parameters; $\varepsilon_i^\tau$ and $\varepsilon_i^{(2)}=\varepsilon_i^e+\varepsilon_i^\mu$. 3)  $T\sim M_1<10^9$ GeV, when all the flavors $(e,\mu,\tau)$ are in equilibrium and distinguishable: fully flavored leptogenesis.\\

Note that the flavor sum on $\alpha$ leads to a vanishing value of the second term in Eq.\eqref{asymp}, since
\begin{equation}
\sum\limits_{\alpha}\text{Im}[h_{ji}(m_D)_{i\alpha}(m_D^*)_{j\alpha}]=\text{Im}[h_{ji}h_{ij}]=\text{Im}|h_{ji}|^2=0,
\end{equation} 
while the first term is proportional to ${\rm Im} (h_{ij}^2)$. Now for both the cases in our model, $h$ has a generic form
\bea
h=\begin{pmatrix}
a^2(1+\kappa_\pm^2)+b_2^2(1+k^2) & ae(1+\kappa_\pm^2)+(1+k^2)b_2c_2 & af(1+\kappa_\pm^2)+(1+k^2)b_2d_2\\ae(1+\kappa_\pm^2)+(1+k^2)b_2c_2 & e^2(1+\kappa_\pm^2)+c_2^2(1+k^2) & ef(1+\kappa_\pm^2)+(1+k^2)c_2d_2\\af(1+\kappa_\pm^2)+(1+k^2)b_2d_2 & ef(1+\kappa_\pm^2)+(1+k^2)c_2d_2 & f^2(1+\kappa_\pm^2)+d_2^2(1+k^2)
\end{pmatrix}\nonumber\\\label{hces}
\eea
with `$\pm$' sign in $\kappa$ are for Case-I and Case-II respectively. Note that the matrix $h$ in Eq.\eqref{hces} is real. Therefore, unflavored leptogenesis which is relevant for the high temperature regime does not take place for any $N_i$  in this model. As mentioned earlier in this section, in general any initial asymmetry produced by the heavier RH neutrinos ($N_{2,3}$) get washed out by lepton number violating $N_1$ related interaction\cite{bari} unless some fine tuned conditions  as discussed in the Sec.\ref{s6} are satisfied. Thus with the assumption that only the decay of $N_1$ matters in generating the CP asymmetry, $\varepsilon_1$ is the relevant quantity for unflavored leptogenesis, but it vanishes in this model. \\

Next, we concentrate on computing the $\alpha$-flavored CP asymmetry in terms of $x_{12}$, $x_{13}$ and the elements of $m_D$. These are necessary ingredients for the fully flavored and the $\tau$-flavored regimes. We find a vanishing value\footnote{This is also true for ${\rm CP}^{\mu\tau}$\cite{Chen:2016ica,Hagedorn:2016lva} since  $(m_D)_{1e}$,  $(m_D)_{2e}$ and $h$ are all real as in our case.} of $\varepsilon_1^e$ while $\varepsilon_1^{\mu,\tau}$ are calculated as
\begin{equation}
\varepsilon_1^\mu=\zeta[b_2k^2(\chi_1+\chi_2)+b_1(\chi_3+\chi_4)-b_2\chi_5]=-\varepsilon_1^\tau.
\label{ep1}
\end{equation} 
In Eq.\eqref{ep1} the real parameters $\zeta$ and $\chi_i$ ($i=1-5)$ are defined as
\begin{eqnarray}
\zeta=[4\pi v^2(b_1^2+(a^2+b_1^2+b_2^2)k^2+b_2k^4)]^{-1}, \label{zeta}\\
\chi_1=b_2(1+k^2)[c_1c_2A_{12}+d_1d_2A_{13}],\\
\chi_2=c[c_1eA_{12}+d_1f A_{13}],\\
\chi_3=b_2(1+k^2)[c^2_1A_{12}-k^2(c_2^2A_{12}-d_2^2A_{13})+d_1^2A_{13}],\\
\chi_4=-ak^2[c_2eA_{12}+d_2fA_{13}],\label{chi4}\\
\chi_5=(1+k^2)[c_1c_2A_{12}+d_1d_2A_{13}]
\end{eqnarray} where $A_{ij}=g(x_{ij})+(1-x_{ij})^{-1}$.\\

 Now for 
$T\sim M_1<$ $10^9$ GeV regime, $Y_B$ is well approximated with\cite{abada}
\bea
Y_B\simeq-\frac{12}{37g^*}\Big[\varepsilon_i^{e}\eta\Big(\frac{151}{179}\tilde{m}_e\Big) + \varepsilon_i^{\mu}\eta\Big(\frac{344}{537}\tilde{m}_\mu\Big)+\varepsilon_i^{\tau}\eta\Big(\frac{344}{537}\tilde{m}_\tau\Big) \Big]
\label{fflv}\eea where $\tilde{m}_\alpha$ are the wash-out masses, defined as 
\bea
\tilde{m}_\alpha=\frac{|(m_D)_{1\alpha}|^2}{M_1} \hspace{1mm} (\alpha=e,\mu,\tau),
\eea $\eta(\tilde{m}_\alpha)$ is the efficiency factor that accounts for the inverse decay and the lepton number violating scattering processes and  $g^*$ is the number of relativistic degrees of freedom in the thermal bath having a value $g^*\approx 106.75$ in the SM. And for $10^9$ GeV $<T\sim M_1 <10^{12}$ GeV, $Y_B$ is approximated with\cite{abada}
\bea
Y_B\simeq-\frac{12}{37g^*}\Big[\varepsilon_i^{(2)}\eta\Big(\frac{417}{589}\tilde{m}_2\Big) + \varepsilon_i^{\tau}\eta\Big(\frac{390}{589}\tilde{m}_\tau\Big)\Big],\label{tflv}
\eea where  $\tilde{m}_2=\sum\limits_{\alpha=e,\mu}\tilde{m}_\alpha=\tilde{m}_e+\tilde{m}_\mu$ and $\varepsilon_i^{(2)}=\sum\limits_{\alpha=e,\mu}\varepsilon_i^{\alpha}=\varepsilon_i^{e}+\varepsilon_i^{\mu}$.\\
  
 At the end we would like to mention the following: Existing literature such as \cite{chen2,Chen:2016ica,Hagedorn:2016lva} also discussed the phenomena of leptogenesis under the framework of residual CP symmetry. They also pointed out the nonoccurrence of unflavored leptogenesis and only the viability of $\tau-$flavored scenario in case of a preserved residual CP symmetry (in particular ${\rm CP}^{\mu\tau}$) in the neutrino sector. Interestingly, Ref.\cite{chen2,Chen:2016ica}  pointed out $M_1$ to be $\mathcal{O}(10^{11}$ GeV) to produce $Y_B$ in the observed range which is also true for our analysis (see numerical section). However the final analysis in Ref.\cite{chen2,Chen:2016ica} is to some extent different from our analysis. In \cite{chen2,Chen:2016ica}, the authors present the variation of $Y_B$ with a single model parameter for a fixed value of $M_1(5\times 10^{11}$ GeV) and for the best fit values of the oscillation parameters. In our analysis, we stick to the near best fit values of the Yukawa parameters for which $Y_B$ is positive. However, as we shall see in the numerical section, we can only constrain the Yukawa parameters scaled by the RH neutrino masses. Thus for a particular set of scaled parameters we can vary the value of $M_1$ freely and obtain an upper and a lower bound on $M_1$ corresponding to the observed upper and lower bound of $Y_B$. Another point is that in our analysis  the sign of the final $Y_B$ depends upon the primed Yukawa parameters and not on the CP phases. However, Ref.\cite{Hagedorn:2016lva} discusses how in a residual CP scheme the sign of $Y_B$ depends upon the low energy CP phases through a correction to the $m_D$ matrix.

%\textbf{We had earlier identified Im $m^{CES}_D$ as the common source of the origin of a nonzero $\theta_{13}$ and leptonic CP violation. A real $m^{CES}_D$ implies vanishing values for $b_2$, $c_2$ and $d_2$ in which case $\epsilon_1^\mu=-\epsilon_1^\tau$ vanishes identically and, as explained in Ref. [14], so does $\theta_{13}$. However, the reverse statement is not true. One could have a vanishing leptonic CP asymmetry simply by setting b1; 2 to zero in (3.12). But, so long as Im $m^{CES}_D$ is nonzero, e.g. through nonvanishing values of $c_2$ and $d_2$, $\theta_{13}$ need not vanish. Indeed, the leptonic CP asymmetry depends rather sensitively on $b_{1,2}$. We shall elaborate on this later in our numerical discussion.}

\section{Numerical analysis: methodology and discussion} \label{s5} 
In order to assess the viability of our theoretical conjecture and consequent outcomes, we present a numerical analysis in substantial detail for both the viable cases. Our method of analysis and organization are as follows. First, we utilize the ($3\sigma$) values of globally fitted neutrino oscillation data (Table \ref{osc1}), together with an upper bound of 0.23 eV\cite{planck} on the sum of the light neutrino masses arising from PLANCK. To fix the absolute neutrino mass scale we assume $m_{max}\approx\sqrt{|\Delta m_{23}|^2}$ which is in general used in the type-I seesaw like models to be consistent with Davidson-Ibarra bound\cite{Davidson:2002qv}. We also discard the possibility of weak washout scenario $K_\alpha=\tilde{m}_\alpha/10^{-3}<1$ which strongly depends upon the initial conditions and likely to be  disfavored by the current oscillation data\cite{bari2}. We first  constrain the parameter space in terms of the rescaled (primed) parameters defined below. 
\bea
 a\longrightarrow a^\prime=\frac{a}{\sqrt{M_1}},\label{rsc1}
 e\longrightarrow e^\prime=\frac{e}{\sqrt{M_2}}, \nonumber\\
    f\longrightarrow f^\prime=\frac{f}{\sqrt{M_2}},
  b_{1,2}\longrightarrow b_{1,2}^\prime=\frac{b_{1,2}}{\sqrt{M_1}},\nonumber\\
   c_{1,2}\longrightarrow c_{1,2}^\prime=\frac{c_{1,2}}{\sqrt{M_2}},
   d_{1,2}\longrightarrow d_{1,2}^\prime=\frac{d_{1,2}}{\sqrt{M_3}}.\label{rsc2}
 \eea 
Then we explore the predictions of the present model in the context of the $\beta\beta_{0\nu}$ experiments for each of the cases. Finally, in order to estimate the value of $Y_B$ we make use of these constrained parameters with a subtlety. Since  we have only constrained the primed parameters, there remains a freedom of various set of independent choices for the parameters of $m_D$ (unprimed) along with $M_i$, for a given set of primed parameters. Note that for the computation of $Y_B$ we need to feed the unprimed parameters and $M_i$ separately. However, for the entire parameter space of primed parameters, it is impractical to generate the unprimed ones for different values of $M_i$ as one ends up with infinite number of choices. For this, from the entire parameter space of the primed parameters, we have considered only that set of primed parameters which corresponds to a positive value of $Y_B$ (sign of $Y_B$ depends upon the primed parameters) and observables that lie near their  best-fit values as dictated by the oscillation data. Then varying $M_1$, we generate the corresponding unprimed set (parameters of $m_D$). Note that here we take only $M_1$ as the free parameter assuming $M_{i+1}/M_i=10^3$ for $i=1,~2$.  Thus for each value of $M_1$ and corresponding unprimed parameters we obtain the final baryon asymmetry $Y_B$. Since $Y_B$ has an observed upper and lower bound, we get an upper and a lower bound for $M_1$  also. Let's now present the numerical results of our analysis in systematic way.\\

\noindent
\underline{{\it Constraints from oscillation data}}\\

\begin{figure}[H]
\begin{center}
\includegraphics[scale=0.32]{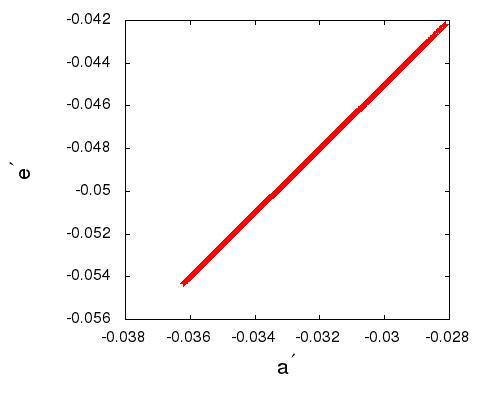}\includegraphics[scale=0.32]{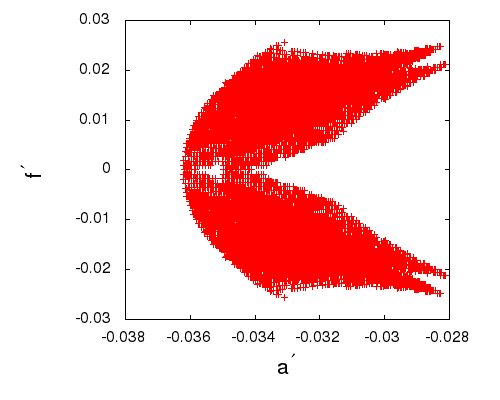}\\
\includegraphics[scale=0.32]{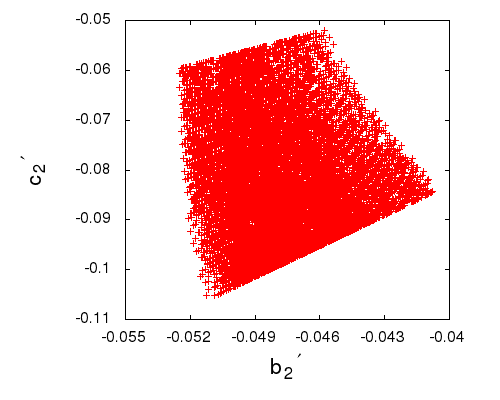}\includegraphics[scale=0.32]{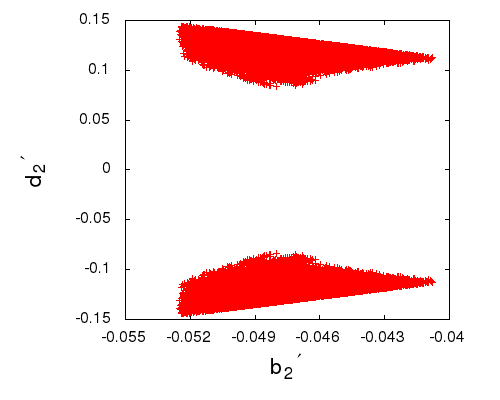}
\caption{ Case-I: Plots of the primed parameters for a normal mass hierarchy.}\label{fg1}
\end{center}
\end{figure}

\begin{figure}[H]
\begin{center}
\includegraphics[scale=0.32]{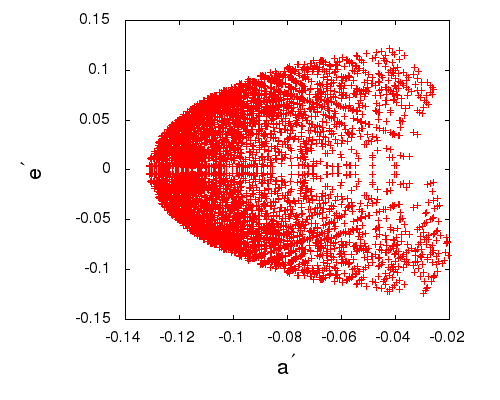}\includegraphics[scale=0.32]{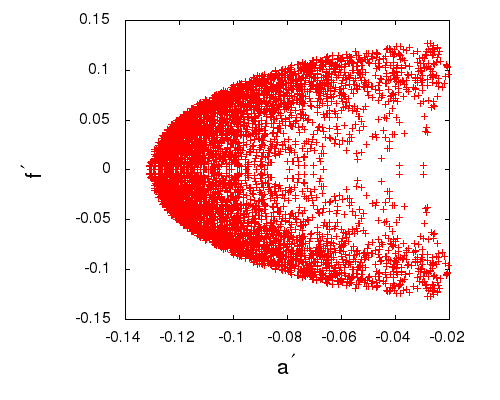}\\
\includegraphics[scale=0.32]{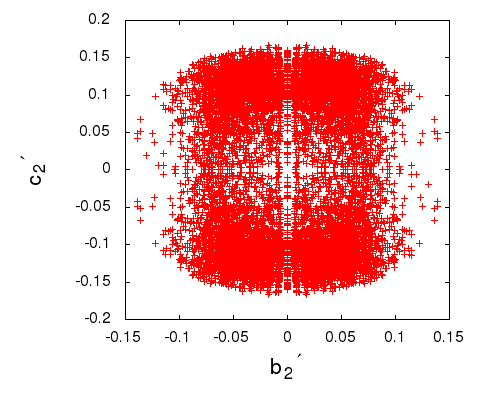}\includegraphics[scale=0.32]{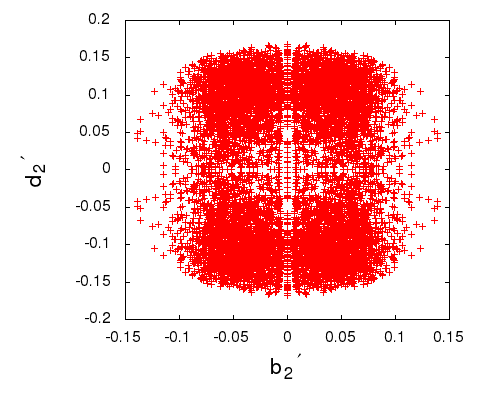}
\caption{ Case-I: Plots of the primed parameters for a inverted mass hierarchy.}\label{fg2}
\end{center}
\end{figure}

\begin{figure}[H]
\begin{center}
\includegraphics[scale=0.32]{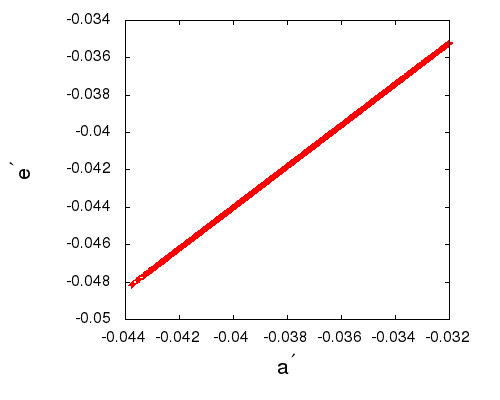}\includegraphics[scale=0.32]{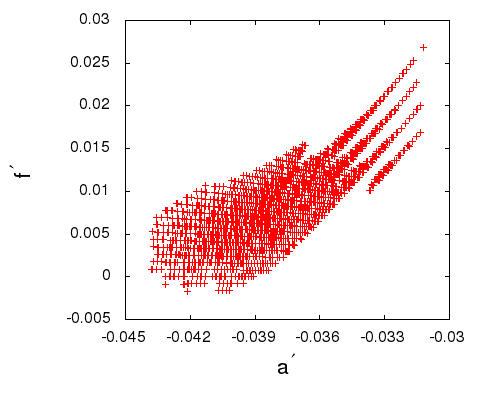}\\
\includegraphics[scale=0.32]{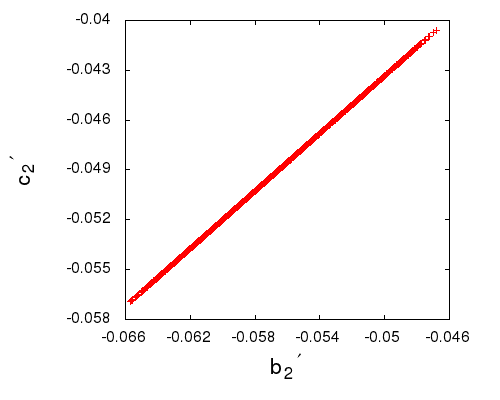}\includegraphics[scale=0.32]{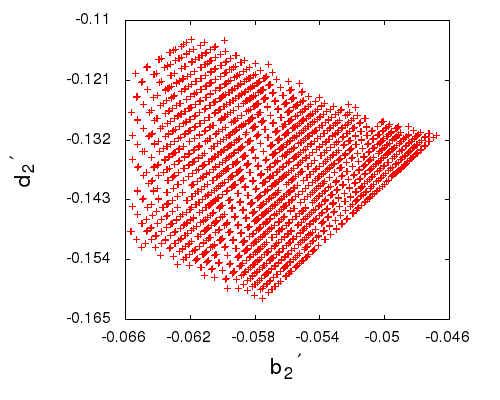}
\caption{ Case-II: Plots of the primed parameters for a normal mass hierarchy.}\label{fg3}
\end{center}
\end{figure}

\begin{figure}[H]
\begin{center}
\includegraphics[scale=0.32]{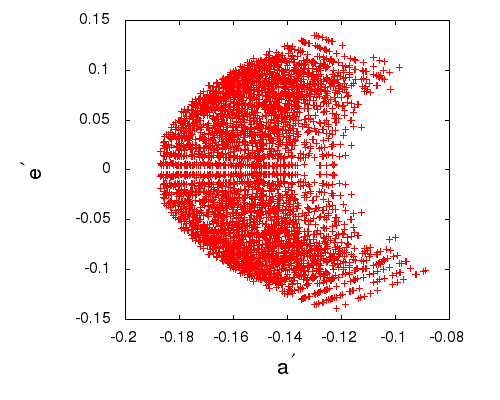}\includegraphics[scale=0.32]{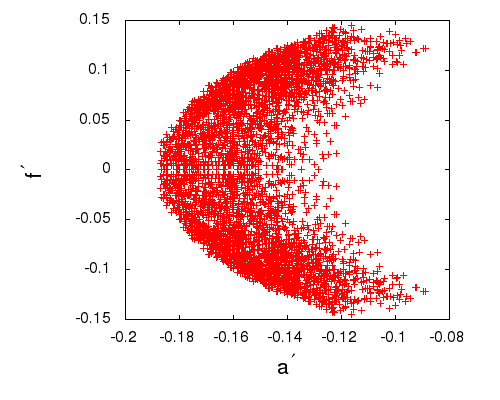}\\
\includegraphics[scale=0.32]{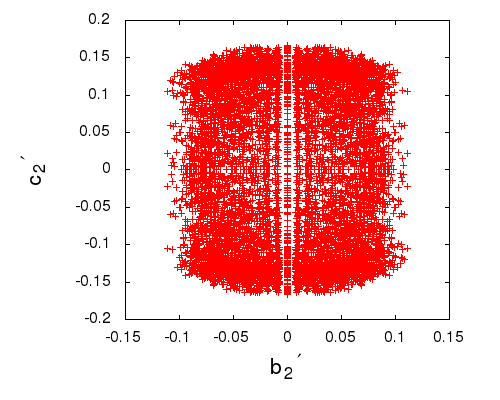}\includegraphics[scale=0.32]{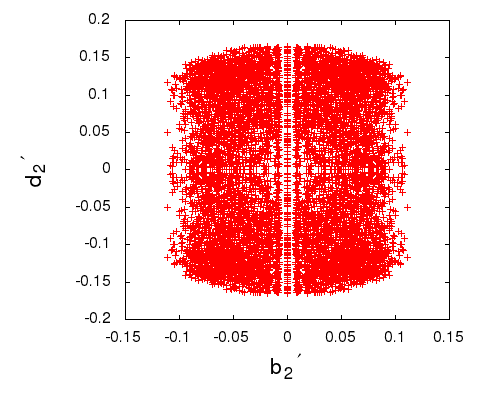}
\caption{ Case-II: Plots of the primed parameters for a normal mass hierarchy.}\label{fg4}
\end{center}
\end{figure}

For each of the viable cases, both the normal ordering (NO) and inverted ordering (IO) of light neutrino masses are found to be permitted over a respectable size of parameter space consistent with the aforementioned experimental constraints. This is interesting since the ordinary SSA predicts $m_3=0$, and thus, inverted light neutrino mass ordering (see Sec.\ref{s2}). However in the extended case both the mass orderings are allowed due to the fact that the matrices $M_\nu^{MS1}$ and $M_\nu^{MS2}$ have nonzero determinant. The ranges of the primed parameters for both the cases I and II are graphically shown in Fig.\ref{fg1}-\ref{fg4}. These plots are basically two dimensional projection of a coupled six dimensional parameter space. In order to constrain the parameter space, the explicit analytic relations that have been implemented in the computer program can be found in Ref.\cite{Adhikary:2013bma} which discusses explicit expressions for the masses and mixing angles for a general $3\times 3$ Majorana mass matrix. \\

 In both the cases,  reduction in the number of parameters upon rescaling led to a constrained range for each of the light neutrino masses as depicted in Table \ref{t3}. It has been found that all the light neutrino mass spectrum are hierarchical. Interestingly, though the upper bound on $\Sigma_im_i$  is fed in as an input constraint, the bound has not been reached up in our model irrespective of the mass ordering. The predictions on $\Sigma_im_i$ are tabulated in Table \ref{t3} for each of the cases.\\
\begin{table}[H]
\begin{center}
\caption{Input values fed into the analysis\cite{osc}.} \label{osc1}
 \begin{tabular}{|c|c|c|c|c|c|} 
\hline 
${\rm Parameters}$&$\theta_{12}$&$\theta_{23}$ &$\theta_{13}$ &$ \Delta m_{21}^2$&$|\Delta m_{31}^2|$ \\
$ $&$\rm (degrees)$ &$\rm (degrees)$ &$ \rm (degrees)$&$10^{-5} \rm (eV^2)$&$ 10^{-3} \rm (eV^2) $ \\
\hline
$3\sigma\hspace{1mm}{\rm ranges/\hspace{1mm}others\hspace{1mm}}$&$31.29-35.91$&$38.3-53.3$&$7.87-9.11$&$7.02-8.09$&$2.32-2.59$\\
\hline
${\rm Best\hspace{1mm}{\rm fit\hspace{1mm}}values\hspace{1mm}(NO)}$&$33.48$&$42.3$&$8.50$&$7.50$&$2.46$\\
\hline
${\rm Best\hspace{1mm}{\rm fit\hspace{1mm}}values\hspace{1mm}(IO)}$&$33.48$&$49.5$&$8.51$&$7.50$&$2.45$\\
\hline
\end{tabular} 
\end{center} 
\end{table}

%\noindent  

\begin{table}[H]
\begin{center}
\caption{Predictions on the light neutrino masses and $\sum_{i}m_i$.} \label{t3}
 \begin{tabular}{|c|c|c|c|c|c|}
 \hline
 \multicolumn{6}{|c|}{\cellcolor{gray!30}{Case-I}} \\
 \hline 
  \multicolumn{3}{|c|}{{Normal Ordering}} & \multicolumn{3}{c|}{{Inverted Ordering}} \\
\hline 
$m_1/10^{-3}$&$m_2/10^{-3}$ &$m_3/10^{-3}$ &$ m_1/10^{-3}$&$m_2/10^{-3}$&$m_3/10^{-3}$ \\
$\rm (eV)$&$ \rm (eV)$ &$\rm (eV)$ &$\rm (eV)$&$ \rm (eV)$&$\rm (eV)$\\
\hline
$4.0-8.5$&$9.28-12.0$&$49-52$&$47-61$&$49-62$&$9-36$\\
\hline
  \multicolumn{3}{|c|}{{$\sum_{i}m_i<0.08$ eV}} & \multicolumn{3}{c|}{{$\sum_{i}m_i<0.16$ eV}} \\
% \hline 
 \hline
 \multicolumn{6}{|c|}{\cellcolor{gray!30}{Case-II}} \\
 \hline 
  \multicolumn{3}{|c|}{{Normal Ordering}} & \multicolumn{3}{c|}{{Inverted Ordering}} \\
\hline 
$m_1/10^{-3}$&$m_2/10^{-3}$ &$m_3/10^{-3}$ &$ m_1/10^{-3}$&$m_2/10^{-3}$&$m_3/10^{-3}$ \\
$\rm (eV)$&$ \rm (eV)$ &$\rm (eV)$ &$\rm (eV)$&$ \rm (eV)$&$\rm (eV)$\\
\hline
$4.1-8.8$&$9.23-13.1$&$48-52$&$47-60$&$49-61$&$10-38$\\
\hline
  \multicolumn{3}{|c|}{{$\sum_{i}m_i<0.08$ eV}} & \multicolumn{3}{c|}{{$\sum_{i}m_i<0.16$ eV}} \\
 \hline 
\end{tabular}  
\end{center} 
\end{table}
\noindent
\underline{{\it Neutrinoless double beta decay ($\beta \beta_{0\nu}$)}}\\ 

\begin{figure}[H]
\begin{center}
\includegraphics[scale=0.2]{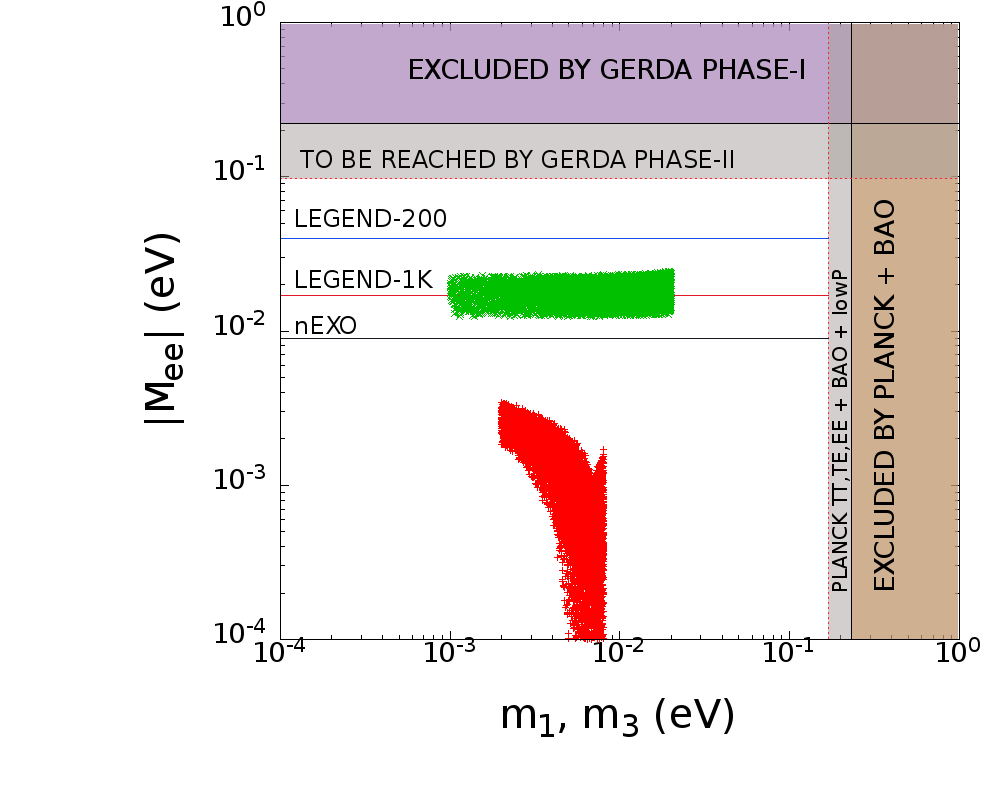}\includegraphics[scale=0.2]{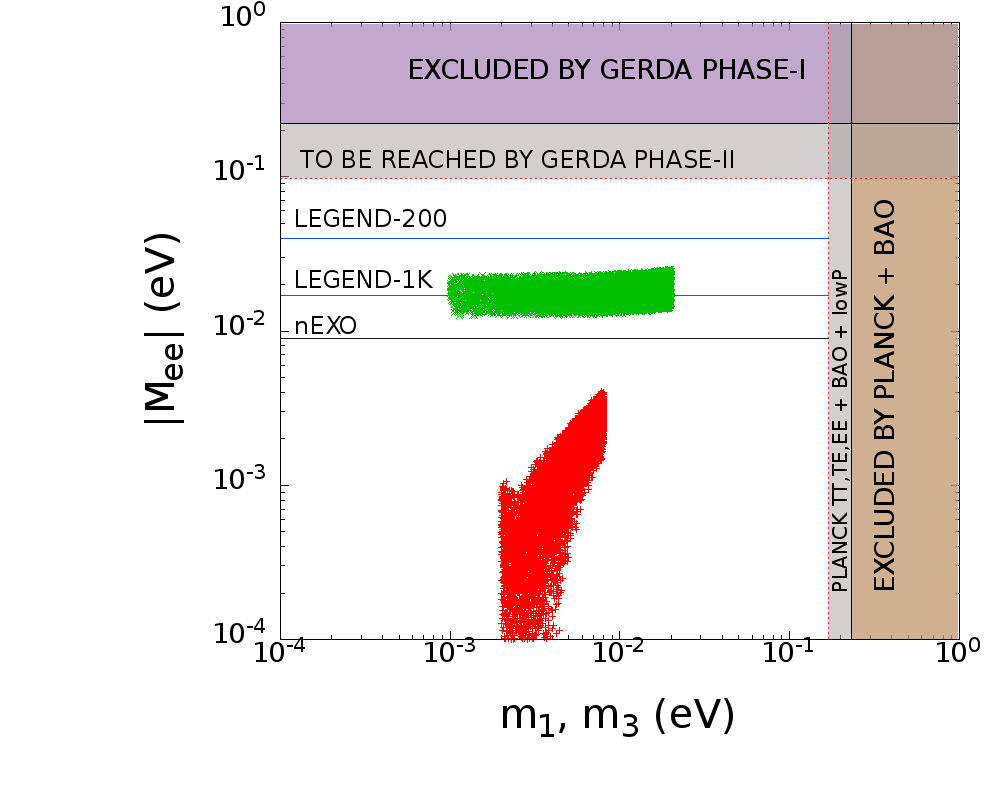}\\
\includegraphics[scale=0.2]{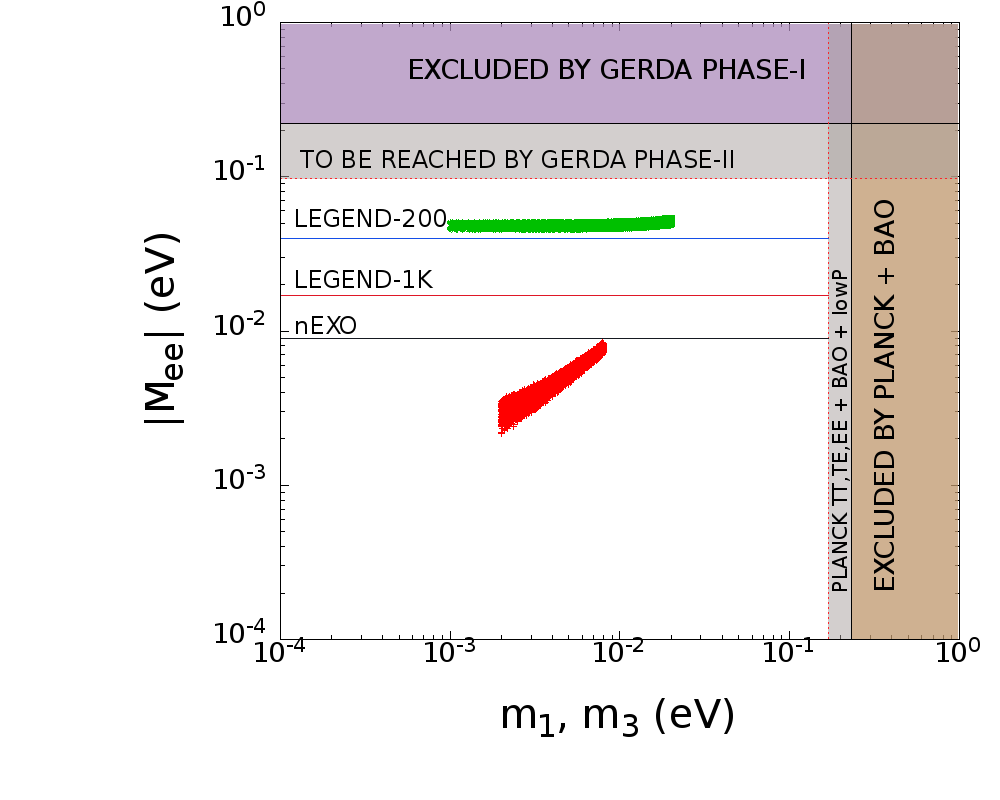}\includegraphics[scale=0.2]{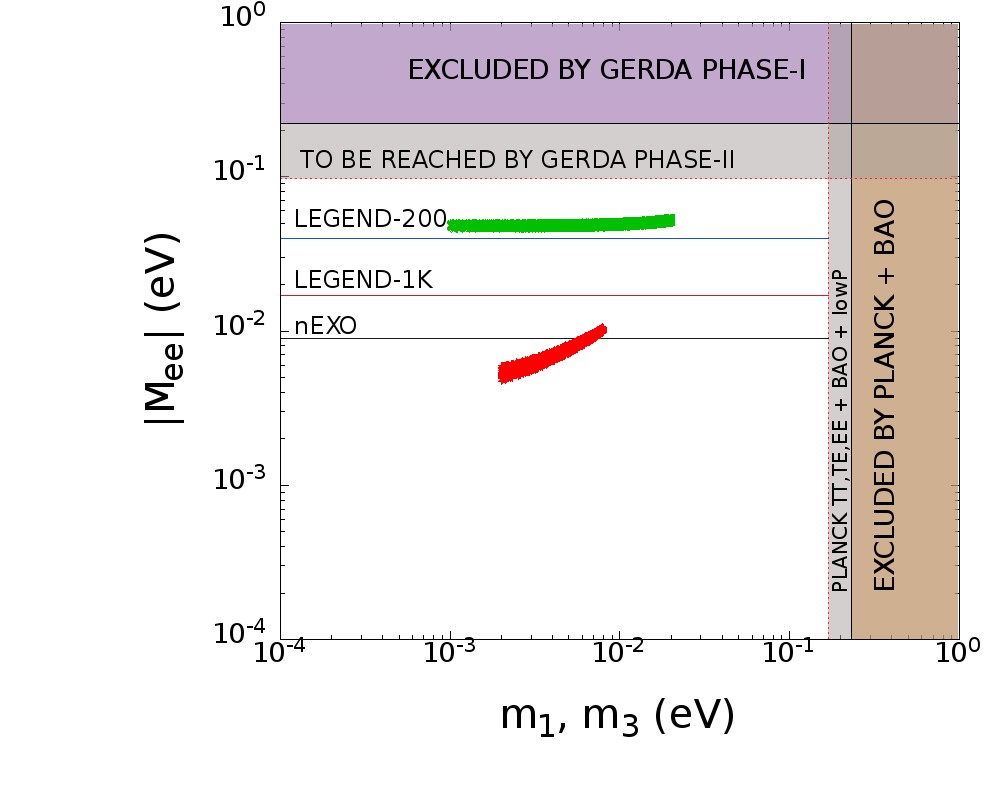}
\caption{Plot of  $|M_{ee}|$ vs. the lightest neutrino mass: the top two figures  represent Case A: $\alpha=\pi$, $\beta=0$ (left) and Case B: $\alpha=\pi$, $\beta=\pi$ (right) while the figures in the lower panel represent Case C: $\alpha=0$, $\beta=0$ (left) and Case D: $\alpha=0$, $\beta=\pi$ (right).}\label{fg5}
\end{center}
\end{figure}

This is a process arising from the decay of a nucleus as
\bea
(A,Z)\longrightarrow (A, Z+2)+2e^-
\eea where the lepton number is violated by 2 units due to the absence of any final state neutrinos. Observation of such decay will lead to the confirmation of the Majorana nature of the neutrinos. The half-life\cite{beta} corresponding to the above decay is given by
\bea
\frac{1}{T^{0\nu}_{1/2}}=G|\mathcal{M}|^2 |M_{ee}|^2m_e^{-2}, 
\eea 
where $G$ is the two-body phase space factor, $\mathcal{M}$ is the nuclear matrix element (NME), $m_e$ is the mass of the electron and  $M_{ee}$ is the (1,1) element of the effective light neutrino mass matrix $M_\nu$. Using the PDG parametrization convention for $U_{PMNS}$\cite{pdg}, the $M_{ee}$ can be written as
\bea
M_{ee}=c_{12}^2c_{13}^2m_1+s_{12}^2c_{13}^2m_2e^{i\alpha}+s_{13}^2m_3e^{i(\beta-2\delta)}.
\eea
Significant upper limits on $|M_{ee}|$ are available from several ongoing experiments. Experiments such as KamLAND-Zen \cite{kam} and EXO \cite{exo} have constrained this value to be $<0.35$ eV. However, till date the most impressive upper bound of 0.22 eV on $|M_{ee}|$ is provided by GERDA phase-I data \cite{gerda1} which is likely to be lowered even further by GERDA phase -II data\cite{gerda2} to around 0.098 eV.
As shown in Ref.\cite{Rome}, existence of $G_3^{(k)}$ in the neutrino mass matrix leads to  four sets of values of the CP-violating Majorana phases $\alpha$ and $\beta$ for each neutrino mass ordering.  Since $|M_{ee}|$ is sensitive to  these phases, we get four different plots for each mass ordering. In Fig.\ref{fg5} we present the plots of $|M_{ee}|$ vs. the lightest neutrino mass ($m_{1,3}$) for both the mass orderings in Case-I only. Apart from  slight changes in the upper and lower limits on $m_{1,3}$, Case-II also leads to  similar plots since it also predicts same results on CP violating phases (i.e.  $\cos\delta=0,\alpha,\beta=0~{\rm or}~\pi$). \\

\noindent

This is evident from Fig.\ref{fg5} that  $|M_{ee}|$ in each plot leads to an upper limit which is below the  reach of the GERDA phase-II data. However, predictions of our model could  be probed by GERDA + MAJORANA experiments \cite{majo}. Sensitivity reach of  other promising experiments such as LEGEND-200 (40 meV), LEGEND-1K (17 meV) and nEXO (9 meV)\cite{Agostini:2017jim} are also shown in Fig.\ref{fg5}. Note that for each case, the entire parameter space corresponding to the inverted mass ordering   could be ruled out by the nEXO reach.  One can also explain the nature of the plots analytically. Let us first consider the inverted mass ordering. In this case, with the approximations $m_3\simeq 0$ and $m_1 \simeq m_2$,  $|M_{ee}|$  simplifies to
\bea
|M_{ee}|=\sqrt{|\Delta m_{32}|^2}c_{13}^2 [\lbrace1-s_{12}^2(1-\cos \alpha)\rbrace^2+s_{12}^4\sin^2\alpha]^{1/2}.\label{ew}
\eea
 Clearly, $|M_{ee}|$ is  not sensitive to the  phases $\beta$ and $\delta$. On the other hand, for $\alpha=\pi$ and $0$ (\ref{ew}) further simplifies to 
 \bea
|M_{ee}|=\sqrt{|\Delta m_{32}|^2}c_{13}^2 [\lbrace 1-2s_{12}^2\rbrace ^2 ]
\eea
and
\bea
|M_{ee}| = \sqrt{|\Delta m_{32}|^2}c_{13}^2
\eea
respectively. Therefore, for $\alpha=\pi$ (cases A, B), $|M_{ee}|$ is suppressed as compared to the case $\alpha=0$ (cases C, D). Now for a normal mass ordering, in addition to the $s_{13}$ suppression, there is a significant interference between the first two terms. If $\alpha=0$, the first two terms interfere constructively  and  we obtain a lower bound ($\sim 10^{-3}$ eV for Case C and $\sim 5\times 10^{-3}$ eV for Case D) despite it being a case of  normal mass ordering of the light neutrinos. This is one of the crucial results of the present analysis. On the other hand, for $\alpha=\pi$, the first two terms interfere destructively and thus a sizable cancellation between them  brings down the value of $|M_{ee}|$ and results in the kinks that is depicted in the lower curves in the top two figures.\\

\noindent
\underline{{\it Baryogenesis via flavored leptogenesis}}\\
\paragraph{}
As mentioned in the beginning of the numerical section, to get a positive $Y_B$, we were obliged to use those value of the primed parameters for which the low energy neutrino parameters predicted from our model lie close to their best fit values dictated by the oscillation experiment. To facilitate this purpose, we define a variable $\chi^2$ in Eq.\eqref{chi2} that measures the deviation of the parameters from their best fit values.  
\begin{equation}
\chi^2= \sum\limits_{i=1}^5 \Big[ \frac{\mathcal{O}_i(th)-\mathcal{O}_i(bf)}{\Delta \mathcal{O}_i} \Big]^2. \label{chi2}
\end{equation}
In Eq.\eqref{chi2}) $\mathcal{O}_i$ denotes the $i^{th}$ neutrino oscillation observable among  $\Delta m^2_{21},\Delta m^2_{32},\theta_{12},\theta_{23}$ and $\theta_{13}$ and the summation runs over all of them. The parenthetical  $th$ stands for  the numerical value of the observable given by our model, whereas $bf$ denotes the best fit value (cf. Table \ref{osc1}). $\Delta \mathcal{O}_i$ in the denominator stands for the measured $1\sigma$ range of $\mathcal{O}_i$. For numerical computation, we choose $M_{i+1}/M_i=10^3\hspace{1mm}(i=1,2)$\footnote{In the next section a detailed discussion is given regarding the  sensitivity of $Y_B$ to the chosen hierarchy of $M_i$.}. First we calculate $\chi^2$  as a function of the primed parameters in their constrained range. For a fixed value of $M_1$, we then start with the minimum value of $\chi^2$ and  we keep on increasing it until $Y_B$ attains a positive value. For that particular $\chi^2$ i.e., for a particular set of primed parameters, we are then able to generate a large set of unprimed parameters by varying $M_1$ over a wide range and can calculate $Y_B$ for each value of $M_1$.  Let's discuss our results case by case for each mass ordering.\\

% $\bullet$ In the temperature regime, $M_1<10^9$ GeV, all the lepton flavours are distinguishable. In this regime, one needs to individually evaluate all $\epsilon^i_\alpha$ i.e., for $\alpha=e,\mu,tau$. Which formula do you use in this case to determine $Y_B$? $\epsilon_{\alpha\alpha}$ is basically $\epsilon_\alpha$ and it is assumed there that $i=1$. So the Davidson, Nir formula is the simplified formula. How the $\epsilon^i_\alpha$ is related to $Y_B$ in general? I don't know it because Dadidson didn't use it. From that formula reduce it to the formulas you use. What are the washout parameters? Any clash with your references? What do they tell you?

\noindent
\textbf{Case-I: $Y_B$ for normal mass ordering of light neutrinos:}\\
\paragraph{$\bf{M_{1}<{10}^{9}}$ GeV:} In this regime, all three lepton flavors $(e,\mu,\tau)$ are distinguishable. Since $\varepsilon_1^{e}=0$, we need to individually evaluate $\varepsilon_1^{\mu,\tau}$ only. Numerically, the maximum value of $|\varepsilon_1^{\mu,\tau}|$ is found to be $\sim 10^{-8}$. $Y_B$ in the observed range cannot be generated with such a small CP asymmetry parameter. Theoretically, this can be understood as an interplay between various quantities. A unique feature in the present model is that the nonzero value of $\theta_{13}$ and $\varepsilon_i$ originated from the imaginary part of the $m_D$ matrix.

\begin{figure}[H]
\begin{center}
\includegraphics[scale=0.4]{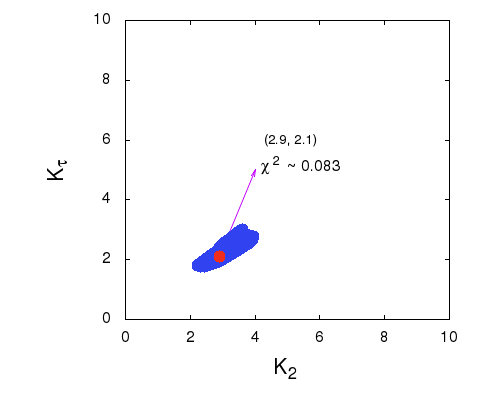}\includegraphics[scale=0.4]{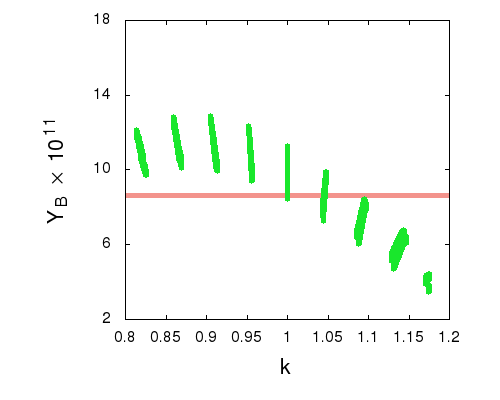}\\
\caption{ The plot on the left hand side shows the range of the wash-out parameters. The red dot corresponds to the minimum value of $\chi^2$ for which a set of primed parameters has been taken to compute $Y_B$. The plot on the right hand side shows a variation of $Y_B$ vs $k$. The red band in the same plot indicates the observed range of $Y_B$.}\label{fg6}
\end{center}
\end{figure}

\noindent
\paragraph{$\bf{{10}^{9}\,\,{\rm{\bf{{\rm GeV}}}}<M_{1}<{10}^{12}}$ GeV:} Before calculating final $Y_B$, we have to look first at the wash-out parameters $K_\alpha=\tilde{m}_\alpha/10^{-3}$ relevant to this mass regime. Since in this regime only $\tau$ flavor is distinguishable, there are two wash-out parameters, $K_\tau$ and $K_2=K_e+K_\mu$. As shown in the first plot of Fig.\ref{fg6}, the entire range of these parameters is not much greater than 1 for the observed range of $Y_B$. Thus the  efficiency factor in Eq.\eqref{tflv}  can be  written for this mild wash-out scenario\cite{abada} as

\bea
\eta(\tilde{m}_\alpha)=\Big[\Big(\frac{\tilde{m}_\alpha}{8.25\times10^{-3}}\Big)^{-1}+\Big(\frac{0.2\times10^{-3}}{\tilde{m}_\alpha}\Big)^{-1.16}\Big]^{-1}.
\eea

We then perform a $\chi^2$ scanning of the primed parameters. It has been found that for $\chi^2_{min}=0.083$ one can have $Y_B$ positive. Basically, In our scheme, Eq.\eqref{tflv} of the present manuscript can be written as
\bea
Y_B\simeq \frac{12}{37g^*} \varepsilon_1^{\mu}\Big[ \eta\Big(\frac{390}{589}\tilde{m}_\tau\Big)-\eta\Big(\frac{417}{589}\tilde{m}_2\Big) \Big].\label{YBsign}
\eea
Thus the sign of $Y_B$ depends upon the sign of $\varepsilon_1^{\mu}$ and the sign of the bracketed quantity. Now from the  primed parameter space we take a particular set, calculate the corresponding $\chi^2$ and then compute $Y_B$. This has been seen that  data sets corresponding to  $\chi^2<0.083$ cannot produce positive $Y_B$, since for those data sets, we get positive values of $\varepsilon_1^{\mu}$ but negative values for the bracketed quantity. A complete data set of the primed parameters and corresponding values of the observables are tabulated in Table \ref{g2nrchi} for $\chi^2_{min}=0.083$. The other parameters i.e., $b_{1},c_1,d_1$ can be calculated using Eq.\eqref{ss1}-\eqref{ss3}.

\begin{table}[H]
\caption{Parameters and observables corresponding  $\chi^2=0.083$ for normal mass ordering.}
\label{g2nrchi}
\begin{center}
 \begin{tabular}{ |c|c|c|c|c|c|c| } 
 \hline
$a^\prime$ & $e^\prime$ & $f^\prime$ & $b_2^\prime$ & $c_2^\prime$ & $d_2^\prime$ & $\chi^2$ \\ \hline
 $-0.036$ & $-0.050$ & $0.003$ & $-0.052$ & $-0.059$ & $-0.122$ & $0.083$\\ \hline
% cell7 & cell8 & cell9 & & &\\ \hline
\hline
\hline
\multicolumn{2}{|c|}{observables} & $\theta_{13}$ & $\theta_{12}$ & $\theta_{23}$ & $\Delta m_{21}^2\times 10^5$ &$|\Delta m_{31}|^2 \times 10^3$\\
\hline
\multicolumn{2}{|c|}{$\chi^2=0.083$} & $8.42^0$ & $33.04^0$ & $42.54^0$ & $7.57~ {\rm (eV)}^2$ &$2.55~{\rm (eV)}^2$\\
\hline
\end{tabular}
\end{center}
\end{table}

\begin{figure}[H]
\begin{center}
\includegraphics[scale=0.4]{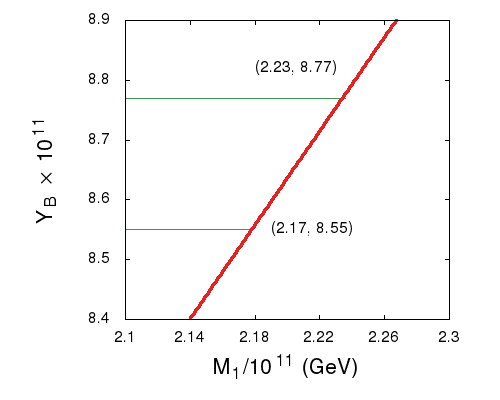}
\caption{A plot of the final $Y_B$ for different values of $M_1$ for a normal light neutrino mass ordering.}
\label{fg7}
\end{center}
\end{figure}

Finally, given the primed data set for that $\chi^2_{min}$, $M_1$  is varied widely to have $Y_B$ in the observed range. For each value of $M_1$, a set of values of the unprimed parameters $\{a,e,f,b_1,c_1,d_1,b_2,c_2,d_2 \}$ is generated. Final $Y_B$ is then calculated for each values of $M_1$ and the corresponding unprimed set. A careful surveillance of the plot in  Fig.\ref{fg7} leads to the conclusion that we can obtain an upper and a lower bound on $M_1$ due to the observed constraint on $Y_B$. In order to appreciate  this  fact  more clearly,   two straight lines have been drawn parallel to the abscissa in the mentioned plot: one at $Y_B=8.55\times10^{-11}$ and the other at $Y_B=8.77\times10^{-11}$.  The values of $M_1$, where the straight lines meet the $Y_B$ vs $M_1$ curve, yield the allowed lower and upper bounds on $M_1$, namely $(M_1)_{lower}=2.17\times10^{11}$ GeV and  $(M_1)_{upper}=2.23\times10^{11}$ GeV.  To explain this linear correlation between $M_1$ and $Y_B$ one could see the expression for $\varepsilon_1^{\mu}$ in Eq.\eqref{epsim}. As we see from Eq.\eqref{epsim}, $\varepsilon_1^\alpha$ is composed of two terms. The first term is proportional to  $M_1/M_j$  while the second term is proportional $(M_1/M_j)^{2}$. Now for the assumed hierarchical scenario ($M_3\gg M_2\gg M_1$), the first term dominates (cf. Eq.\ref{epssimp}) and effectively $\varepsilon_1^\alpha$ becomes proportional to $M_1$ (theoretically which is not the case due to the presence of the second term). Now in Eq.\eqref{tflv}, in the expression of $Y_B$, the wash-out parameters only depend upon the primed parameters. Thus effectively the final baryon asymmetry $Y_B$ is also proportional to $M_1$. One might also ask about the narrow range for $M_1$ as we see in the Fig.\ref{fg7}. Basically we have presented our result for a particular set of  primed parameters (for $\chi^2_{min}=0.083$). In principle one can take the entire primed parameter space of our model and compute the corresponding results on $Y_B$ and $M_1$ for each set of primed parameters. In that case (for the entire parameter space) the range of $M_1$ should not be as narrow as we see in this case.

%\noindent
From Table \ref{g2nrchi}, we infer that  $\theta_{23}=42.54^0$ corresponding to $\chi^2_{min}=0.083$. Since theoretically $\theta_{23}$ is related only with a single model parameter $k$ (cf. Eq.\eqref{th23k}) and unlike the other parameters of $m_D$ (discussed earlier in this section) value of $k$ does not depend upon the variation of $M_1$, $\theta_{23}$ remain fixed for the entire range of $M_1$ that corresponds to the observed range of $Y_B$. Thus an experimentally appealing conclusion of this scheme is that, given the observed range of $Y_B$, the octant of $\theta_{23}$ is determined ($<45^0$). One can also check the sensitivity of the produced $Y_B$ to the entire range of $\theta_{23}$ in a slightly different way.  It is trivial to find out the analytic form of $Y_B$ that explicitly depend upon $\theta_{23}$, by replacing $k=(\tan\theta_{23})^{-1}$ in the expression of $\varepsilon_1^\mu$ and $m_\alpha$ in Eq.\eqref{YBsign}. Thus for a fixed value of $M_1$ one can use the entire  parameter space of the primed parameters and $k$ to compute the final $Y_B$. From the plot on the right panel of Fig.\ref{fg6}, we see that the value of $k$ is always greater that 1 for $Y_B$ to be in the observed range (represented by the red narrow strip in Fig.\ref{fg6}). This is certainly for a particular value of $M_1(6.79\times10^{11}{\rm GeV})$. As previously mentioned, $Y_B$ is almost proportional to $M_1$, thus lowering the value of the latter below $6.79\times10^{11}{\rm GeV}$ would cause a downward movement of the overall pattern of the $Y_B$ vs. $k$ plot in Fig.\ref{fg6}. Thus for the observed range of $Y_B$, along with the values $k>1$, there would be other values of $k$ which are less than one. It is seen that for the normal mass ordering in Case-II a similar lower limit on $M_1$ exist that dictates the octant of $\theta_{23}$ for the the observed range of $Y_B$.\\

\begin{figure}[H]
\begin{center}
\includegraphics[scale=0.4]{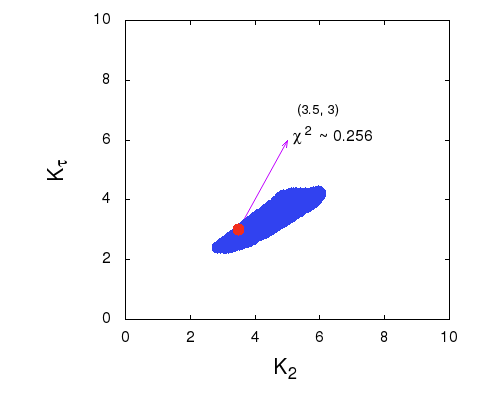}\includegraphics[scale=0.4]{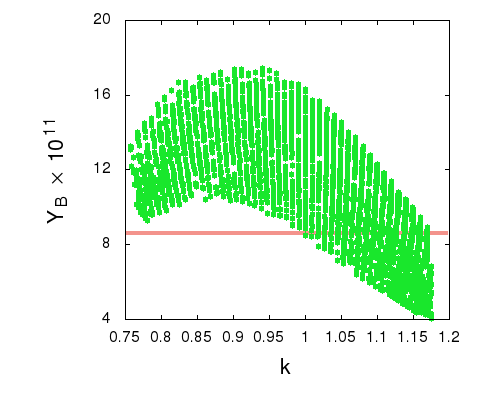}\\
\caption{ The plot on the left hand side shows the range of the wash-out parameters. The red dot corresponds to the minimum value of $\chi^2$ for which a set of primed parameter has been taken to compute $Y_B$. The plot on the right hand side shows a variation of $Y_B$ vs $k$. The red band in the same plot indicates the observed range of $Y_B$.}\label{fg9}
\end{center}
\end{figure}

We would like to stress that the lower bound obtained in the second approach is different from that is obtained in the first one. This is simply because the ways to obtain these bounds are different. In the first approach we take the best fit values of the primed parameters and $k$ and then vary $M_1$ to obtain the observed range of $Y_B$ which in turn leads to an upper and a lower bound on $M_1$. However, in the second approach, we take the entire primed parameter space along with the allowed range for $k$ and then compute $Y_B$ for a fixed value of $M_1$. The $Y_B$ vs. $k$ plot in Fig.\ref{fg6} is for $M_1=6.79\times10^{11}{\rm GeV}$ which represents the lower bound on $M_1$ above which we always get $k>1$ for the observed range of $Y_B$. Now what happens if we further lower the value of $M_1$ from $6.79\times10^{11}{\rm GeV}$ in the second approach? As discussed previously, this would imply a downward movement of $Y_B$ vs $k$ curve in Fig.\ref{fg6} or in Fig.\ref{fg9}. In that case both $k>1$ and $k<1$ values are possible for the observed range of $Y_B$. Obviously this has an impact on the results obtained in the first method. We know from the first method that if we choose the best fit value of $k$, the allowed  range of $M_1$ should be read from Fig.\ref{fg7}. This does not necessarily mean that for this range of $M_1$, other values of $k$ are not possible (obviously those values of $k$ should not be the best fit values then) since the range shown in Fig.\ref{fg7} is below $M_1=6.79\times 10^{11}$ GeV.

\paragraph{$\bf{M_{1}>{10}^{12}}$ GeV:} It has been shown that $Y_B=0$ here for our model.\\

\noindent
\textbf{Case-I: $Y_B$ for inverted mass ordering of light neutrinos:}\\

\noindent
Following the same procedure as for the normal mass ordering, a final discussion for each regime is summarized as follows. 

\paragraph{${\bf{M_{1}<{10}^{9}}}$ GeV:} Similar to the normal ordering, the $|\varepsilon_1^{\mu,\tau}|$ can have values  at most the order of $10^{-8}$ which  is not sufficient to let $Y_B$ come within its observed range.

\begin{figure}[H]
\begin{center}
\includegraphics[scale=0.4]{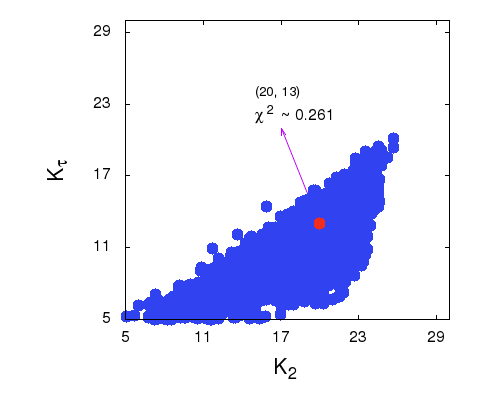}\includegraphics[scale=0.4]{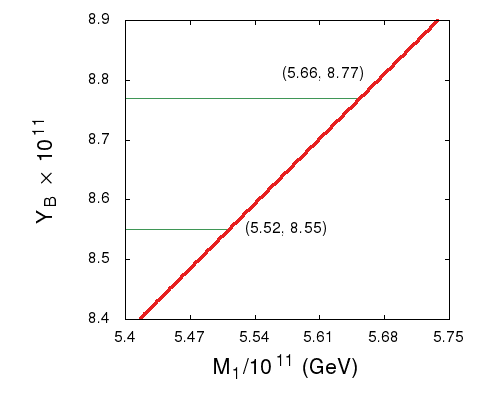}\\
\caption{The plot on the left hand side shows the range of the wash-out parameters. The red dot corresponds to the minimum value of $\chi^2$ for which a set of primed parameter has been taken to compute $Y_B$. The plot on the right hand side  shows final $Y_B$ for different values of $M_1$ for the inverted light neutrino mass ordering.}\label{fg8}
\end{center}
\end{figure}

\paragraph{$\bf{{10}^{9}\,\,{\rm{\bf{{\rm GeV}}}}<M_{1}<{10}^{12}}$ GeV:}Unlike the previous case the ranges of the the wash-out parameters (cf. Fig.\ref{fg8}) favors  a strong wash-out scenario. 

Thus the  efficiency factor in Eq.\eqref{tflv} can be  written for this strong wash-out scenario\cite{abada}as

\bea
\eta(\tilde{m}_\alpha)=\Big[\Big(\frac{0.55\times10^{-3}}{\tilde{m}_\alpha}\Big)^{1.16}\Big].
\eea
For $\chi^2_{min}=0.261$, a set of primed parameters is obtained (cf. Table \ref{invg2chi}). Then similar to the previous case, varying $M_1$ in a wide range, a lower and upper bound on $M_1$, namely $(M_1)_{lower}=5.52\times10^{11}$ GeV and  $(M_1)_{upper}=5.66\times10^{11}$ GeV is obtained for the observed range of $Y_B$. A plot of $Y_B$ vs $M_1$ is shown in the right panel of Fig.\ref{fg8}. 

\noindent
\begin{table}[!ht]
\caption{Parameters and observables corresponding  $\chi^2=0.261$ for inverted hierarchy.}
\label{invg2chi}
\begin{center}
  \begin{tabular}{ |c|c|c|c|c|c|c| } 
 \hline
$a^\prime$ & $e^\prime$ & $f^\prime$ & $b_2^\prime$ & $c_2^\prime$ & $d_2^\prime$ & $\chi^2$ \\ \hline
 $-0.043$ & $-0.065$ & $0.116$ & $0.130$ & $-0.019$ & $0.039$ & $0.261$\\ \hline
% cell7 & cell8 & cell9 & & &\\ \hline
\hline
\hline
\multicolumn{2}{|c|}{observables} & $\theta_{13}$ & $\theta_{12}$ & $\theta_{23}$ & $\Delta m_{21}^2\times 10^5$ &$|\Delta m_{31}|^2 \times 10^3$\\
\hline
\multicolumn{2}{|c|}{$\chi^2=0.261$} & $8.54^0$ & $34.07^0$ & $49.37^0$ & $7.53~ {\rm (eV)}^2$ &$2.40~{\rm (eV)}^2$\\
\hline
\end{tabular}
\end{center}
\end{table}  
 
 \paragraph{$\bf{M_{1}>{10}^{12}}$ GeV:} Once again, $Y_B=0$ in this regime, for the present model.\\

\noindent
\textbf{Case-II: $Y_B$ for normal mass ordering of light neutrinos:}\\

The analysis has been done exactly in the same way as was in the previous case. A systematic presentation of the obtained results is the following.
\paragraph{$\bf{M_{1}<{10}^{9}}$ GeV:} Again, $Y_B$ in the observed range cannot be generated due to the small value of  $|\varepsilon_1^{\mu,\tau}|$.  
\noindent
\paragraph{$\bf{{10}^{9}\,\,{\rm{\bf{{\rm GeV}}}}<M_{1}<{10}^{12}}$ GeV:} Similar to the previous normal hierarchical case, the wash-out parameters here also suggest a mild wash-out scenario (cf. Fig.\ref{fg9}).\\

\begin{figure}[H]
\begin{center}
\includegraphics[scale=0.4]{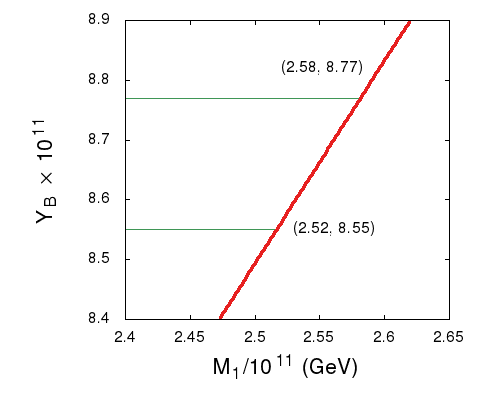}
\caption{A plot of the final $Y_B$ for different values of $M_1$ for the normal light neutrino mass ordering.}
\label{fg10}
\end{center}
\end{figure}

 For $\chi^{2}_{min}=0.256$, a set of rescale parameter has been found and then varying $M_1$ in a wide range, a lower and a upper bound on $M_1$ are obtained as shown in the Fig.\ref{fg10}. Note that in this case also $\theta_{23}<45^0$ (Table \ref{chi_nh}) for the minimum $\chi^2$ that produce $Y_B$ positive and in the observed range. Similar to the case of normal mass ordering in Case-I, here we also show a $Y_B$ vs $k$ plot (cf. Fig.\ref{fg9}) and infer that  there exists a lower limit  $8.2\times 10^{11}{\rm GeV}$ on $M_1$ for which $k>1$, i.e.,  $\theta_{23}<45^0$  for $Y_B$ to be in the observed range.

\begin{table}[H]
\caption{Parameters and observables corresponding  $\chi^2=0.256$ for normal mass ordering.}
\label{chi_nh}
\begin{center}
 \begin{tabular}{ |c|c|c|c|c|c|c| } 
 \hline
$a^\prime$ & $e^\prime$ & $f^\prime$ & $b_2^\prime$ & $c_2^\prime$ & $d_2^\prime$ & $\chi^2$ \\ \hline
 $-0.042$ & $-0.046$ & $-0.005$ & $-0.065$ & $-0.056$ & $-0.128$ & $0.256$\\ \hline
% cell7 & cell8 & cell9 & & &\\ \hline
\hline
\hline
\multicolumn{2}{|c|}{observables} & $\theta_{13}$ & $\theta_{12}$ & $\theta_{23}$ & $\Delta m_{21}^2\times 10^5$ &$|\Delta m_{31}|^2 \times 10^3$\\
\hline
\multicolumn{2}{|c|}{$\chi^2=0.256$} & $8.37^0$ & $33.08^0$ & $43.49^0$ & $7.55~ {\rm (eV)}^2$ &$2.55~{\rm (eV)}^2$\\
\hline
\end{tabular}
\end{center}
\end{table}
\paragraph{$\bf{M_{1}>{10}^{12}}$ GeV:} It has been shown that $Y_B=0$ here for our model.\\

\noindent
\textbf{Case-II: $Y_B$ for inverted mass ordering of light neutrinos:}\\

\noindent
Proceeding  exactly in the same manner as for the normal mass ordering, a brief discussion for each  regime goes as follows. 

\paragraph{${\bf{M_{1}<{10}^{9}}}$ GeV:} Similar to the normal ordering, the $|\varepsilon_1^{\mu,\tau}|$ can have values  at most the order of $10^{-8}$ which  is not sufficient to let $Y_B$ come within its observed range.

\begin{figure}[H]
\begin{center}
\includegraphics[scale=0.4]{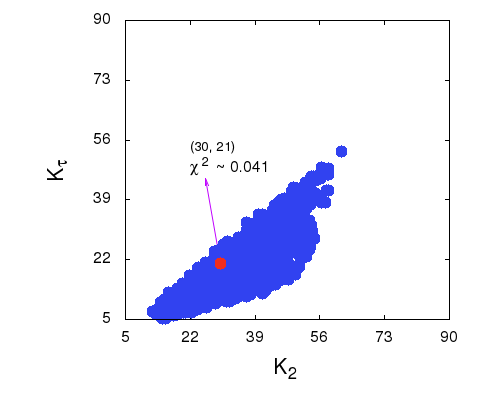}\includegraphics[scale=0.4]{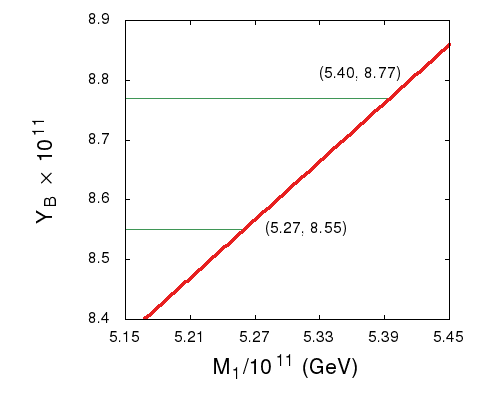}\\
\caption{The plot on the left hand side shows the range of the wash-out parameters. The red dot corresponds to the minimum value of $\chi^2$ for which a set of primed parameter has been taken to compute $Y_B$. The plot on the right hand side  shows final $Y_B$ for different values of $M_1$ for the inverted light neutrino mass ordering.}\label{fg11}
\end{center}
\end{figure}

\paragraph{$\bf{{10}^{9}\,\,{\rm{\bf{{\rm GeV}}}}<M_{1}<{10}^{12}}$ GeV:}Unlike the previous case the ranges of the  wash-out parameters (cf. Fig.\ref{fg11}) favors  a strong wash-out scenario.  For $\chi^2_{min}=0.041$ a set of primed parameters is obtained (cf Table \ref{invg1chi}). Then similar to the previous case varying $M_1$ in a wide range a lower and upper bound on $M_1$, namely $(M_1)_{lower}=5.27\times10^{11}$ GeV and  $(M_1)_{upper}=5.40\times10^{11}$ GeV is obtained for the observed range of $Y_B$. A plot of $Y_B$ vs $M_1$ is shown in the right panel of Fig.\ref{fg11}. 

%\end{center}

\noindent
\begin{table}[!ht]
\caption{Parameters and observables corresponding  $\chi^2=0.041$ for inverted hierarchy.}
\label{invg1chi}
\begin{center}
 \begin{tabular}{ |c|c|c|c|c|c|c| } 
 \hline
$a^\prime$ & $e^\prime$ & $f^\prime$ & $b_2^\prime$ & $c_2^\prime$ & $d_2^\prime$ & $\chi^2$ \\ \hline
 $-0.123$ & $-0.084$ & $0.123$ & $0.104$ & $-0.052$ & $-0.096$ & $0.041$\\ \hline
% cell7 & cell8 & cell9 & & &\\ \hline
\hline
\hline
\multicolumn{2}{|c|}{observables} & $\theta_{13}$ & $\theta_{12}$ & $\theta_{23}$ & $\Delta m_{21}^2\times 10^5$ &$|\Delta m_{31}|^2 \times 10^3$\\
\hline
\multicolumn{2}{|c|}{$\chi^2=0.041$} & $8.71^0$ & $33.43^0$ & $49.23^0$ & $7.58~ {\rm (eV)}^2$ &$2.44~{\rm (eV)}^2$\\
\hline
\end{tabular}
\end{center}
\end{table}  
 
 \paragraph{$\bf{M_{1}>{10}^{12}}$ GeV:} Once again, $Y_B=0$ here for the present model.\\

 A compact presentation of the final conclusions regarding $Y_B$ from the numerical analysis  is given in Table \ref{tf}. 
\begin{table}[H]
\begin{center}
\caption{Final statements on $Y_B$ for different mass regimes. } \label{tf}
 \begin{tabular}{|c|c|c|c|} 
 \cline{1-4}
 \multicolumn{4}{|c|}{\cellcolor{gray!30}Case-I}\\
\cline{1-4}
\multicolumn{1}{|c|}{${\rm Type}$}&${M_1<10^{9}\hspace{1mm}{\rm GeV}}$&$10^{9}\hspace{1mm}{\rm GeV}<M_1<10^{12}\hspace{1mm}{\rm GeV}$ &${M_1>10^{12}\hspace{1mm}{\rm GeV}}$\\
\cline{1-4}
\multicolumn{1}{|c|}{$\pbox{20cm}{{\rm Normal}\\ {\rm Ordering}}$} & $\pbox{20cm}{{\rm Ruled~out~since~$Y_B$ }\\{\rm is~below~the~observed~range}\\{\rm for~any~$\chi^2$.}}$ & $\pbox{20cm}{{\rm $Y_B$ within \hspace{1mm}the~ observed \hspace{1mm} range}\\{\rm for~$\chi^2_{min}=0.083$.}}$ &$\pbox{20cm}{{\rm Ruled~out}\\{\rm since $Y_B=0.$}}$ \\
\cline{1-4}
\multicolumn{1}{|c|}{$\pbox{20cm}{{\rm Inverted}\\ {\rm Ordering}}$}&$\pbox{20cm}{{\rm Ruled~out~since~$Y_B$ }\\{\rm is~below~the~observed~range}\\{\rm for~any~$\chi^2$.}}$&$ \pbox{20cm}{{\rm $Y_B$ within \hspace{1mm}the ~ observed \hspace{1mm} range}\\{\rm for~$\chi^2_{min}=0.261$.}}$&$\pbox{20cm}{{\rm Ruled~out}\\{\rm since $Y_B=0.$}}$\\
\cline{1-4}
 \cline{1-4}
 \multicolumn{4}{|c|}{\cellcolor{gray!30}Case-II}\\
\cline{1-4}
\multicolumn{1}{|c|}{${\rm Type}$}&${M_1<10^{9}\hspace{1mm}{\rm GeV}}$&$10^{9}\hspace{1mm}{\rm GeV}<M_1<10^{12}\hspace{1mm}{\rm GeV}$ &${M_1>10^{12}\hspace{1mm}{\rm GeV}}$\\
\cline{1-4}
\multicolumn{1}{|c|}{$\pbox{20cm}{{\rm Normal}\\ {\rm Ordering}}$} & $\pbox{20cm}{{\rm Ruled~out~since~$Y_B$ }\\{\rm is~below~the~observed~range}\\{\rm for~any~$\chi^2$.}}$ & $\pbox{20cm}{{\rm $Y_B$ within \hspace{1mm}the~ observed \hspace{1mm} range}\\{\rm for~$\chi^2_{min}=0.256$.}}$ &$\pbox{20cm}{{\rm Ruled~out}\\{\rm since $Y_B=0.$}}$ \\
\cline{1-4}
\multicolumn{1}{|c|}{$\pbox{20cm}{{\rm Inverted}\\ {\rm Ordering}}$}&$\pbox{20cm}{{\rm Ruled~out~since~$Y_B$ }\\{\rm is~below~the~observed~range}\\{\rm for~any~$\chi^2$.}}$&$ \pbox{20cm}{{\rm $Y_B$ within \hspace{1mm}the ~ observed \hspace{1mm} range}\\{\rm for~$\chi^2_{min}=0.041$.}}$&$\pbox{20cm}{{\rm Ruled~out}\\{\rm since $Y_B=0.$}}$\\
\cline{1-4}
\end{tabular} 
\end{center} 
\end{table}
Before concluding this section we want to stress the following point. In this model, the imaginary part of $m_D^{MS}$ of Eq.\eqref{mdcs} plays a crucial role. Absence of the latter leads to a vanishing $\theta_{13}$, and thus undetermined value of $\delta$ and most importantly a vanishing value of $\varepsilon^\alpha_i$. Thus the model  addresses a common origin of $\theta_{13}$, CP violation and leptogenesis. However, although the parameters in the imaginary part of $m_D^{MS}$ are correlated with $Y_B$, from Eq.\eqref{ep1} we see the parameter $b_1$ is also very much sensitive to $\varepsilon^\alpha_1$. For example, for $b_2=0$ and $c_2,~d_2\neq 0$,  Eq.\eqref{ep1} is simplified as
 \bea
 \varepsilon_1^\mu= 4\pi v^2[b_1^2+(a^2+b_1^2)k^2]^{-1}b_1\chi_4=-\varepsilon_1^\tau,
 \eea
 where $\chi_4=f(c_2,d_2)$ as defined in Eq.\eqref{chi4}. Now if $b_1$ vanishes $\varepsilon_1^\mu$, hence, $Y_B$ vanishes but due to nonvanshing value of $c_2,d_2$ one obtains $\theta_{13}\neq 0$. However, to obtain a nonzero $Y_B$, along with a nonvanishing  $b_1$, one always needs $\chi_4\neq 0$ which in turn implies a nonzero $\theta_{13}$. Thus in this model a nonzero $\theta_{13}$ does not always imply a nonzero $Y_B$ but the reverse is not true.
\section{Effect of $N_{2,3}$ on $Y_B$}\label{s6}
%Indirect effect:
In our  analysis, the effect of the  two heavier neutrinos ($N_2$, $N_3$) on the produced final baryon asymmetry has been neglected with the assumption that the asymmetries produced by the decays of both of them get  washed out\cite{bari}. In this section, we present a brief discussion on the sensitivity of the heavier neutrinos to final $Y_B$. There are two ways that such a sensitivity might arise as elaborated below.\\

\noindent
\underline{Indirect effect of $N_{2,3}$:}\\

Though the neutrino oscillation data are fitted with the primed parameters, cf Eq.\eqref{rsc1}, for computing  the quantities related to leptogenesis, e.g., $\varepsilon_1^\alpha$, we need to evaluate the  unprimed ones, i.e. the Dirac mass matrix elements. It is interesting to see whether the final baryon asymmetry is affected by the chosen hierarchies of the RH neutrinos. We find that  the final $Y_B$ is not much sensitive to $M_{2,3}$.  One can appreciate this statement  by simplifying the CP asymmetry parameters of Eq.\eqref{asymp} to 
\bea
 \varepsilon_1^{\alpha}=-\frac{3}{8\pi v^2 h_{11}}\sum\limits_{j =2,3} \frac{M_1}{M_j}\hspace{1mm} {\rm Im}[\hspace{1mm}h_{1j}(m_D)_{1 \alpha}(m_D^*)_{j \alpha}]-\frac{1}{4\pi v^2 {h}_{11}}\sum_{j=2,3} \frac{M_1^2}{M_j^2} {\rm Im}[\hspace{1mm}{h}_{j1}(m_D)_{1 \alpha }({m_D}^*)_{ j \alpha }],
%(\varepsilon^\alpha_i)_3&=&\frac{v^2}{H_{ii}}\sum_{j,k}|\epsilon_{ijk}|\frac{(1-x_{ij})\{(1-x_{ik})Im(Z^\alpha_{ijk})
%-\frac{Re(Z^\alpha_{ijk})H_{kk}}{4\pi v^2} \}}
%{\{(1-x_{ij})^2+\delta_{ijk}\}\{(1-x_{ik})^2+}\label{p3}
\label{epsim}
\eea
after approximating $g(x_{1j})$ of Eqs.\eqref{cpa2} to  $g(x_{1j})=-\frac{3}{2\sqrt{x_{1j}}}$ for $x_{1j}\gg 1$. The last term of Eq.\eqref{epsim} is  suppressed because it is of second order in $x_{1j}^{-1}$. Having two parts  for $j=2,3$,  $j=3$ term of the first term of Eq.\eqref{epsim}  has a negligible effect on $\varepsilon_1^\alpha$ since $M_3$ is much larger than $M_1$ and  $f,d_1$ and $d_2$ have values of the order of the other Dirac components. Now for $j=2$, $\varepsilon_1^{\alpha}$ is simplified as 
\bea
\varepsilon_1^{\mu}=-\frac{3M_1}{8\pi v^2h_{11}}[(ae^\prime +b_1 c_1^\prime + b_2 c_2^\prime)(b_2c_1^\prime + b_1 c_2^\prime)]=-\varepsilon_1^{\tau} \label{epssimp}
\eea  
with $\varepsilon_1^{e}=0$ as already shown in Sec.\ref{s4}. Since  the primed parameters are fixed by the oscillation data, $\varepsilon_1^{\mu,\tau}$ are practically insensitive to the value of $M_2$. However, for the numerical computation of the final baryon asymmetry, we take into account each term in Eq.\eqref{epsim} with two different mass hierarchical schemes for the heavy neutrinos, e.g, $M_{i+1}/M_i=10^2$ and $M_{i+1}/M_i=10^4$ where $i$ can take the values 1, 2. Note that in the previous section we have already computed  $Y_B$ for $M_{i+1}/M_i=10^3$. The outcome of the numerical analysis is that though the chosen mass ratios of the RH neutrinos are altered, changes in the lower and upper bounds on $M_1$ are not  significant for the observed range of $Y_B$. For convenience, for each case and light neutrino mass ordering, the variation of $Y_B$ with $M_1$ for different mass ratios has been presented  in Table \ref{t8}.

\begin{table}[H]
\begin{center}
\caption{Lower and upper bounds on $M_1$ for different mass ratios of the RH neutrinos ($i=1,2$).} \label{t8}
 \begin{tabular}{|c|c|c|c|} 
\hline 
 \multicolumn{4}{|c|}{\cellcolor{gray!30}Case-I: Normal light neutrino ordering}\\
${\rm Hierarchies~\rightarrow }$&$M_{i+1}/M_i=10^2$&$M_{i+1}/M_i=10^3$ &$M_{i+1}/M_i=10^4$\\
\hline
${\rm Upper~bound~(GeV)}$&$2.21\times 10^{11}$&$2.23 \times 10^{11}$ &$2.25 \times 10^{11}$ \\
\hline
${\rm Lower~bound~~(GeV)}$&$2.16 \times 10^{11}$&$2.17 \times 10^{11}$&$2.18 \times 10^{11}$\\
\hline
 \multicolumn{4}{|c|}{\cellcolor{gray!30}Case-I: Inverted light neutrino ordering}\\
${\rm Hierarchies~\rightarrow }$&$M_{i+1}/M_i=10^2$&$M_{i+1}/M_i=10^3$ &$M_{i+1}/M_i=10^4$\\
\hline
${\rm Upper~bound~(GeV)}$&$5.64\times 10^{11}$&$5.66 \times 10^{11}$ &$5.67 \times 10^{11}$ \\
\hline
${\rm Lower~bound~~(GeV)}$&$5.51 \times 10^{11}$&$5.52 \times 10^{11}$&$5.54 \times 10^{11}$\\
\hline
 \multicolumn{4}{|c|}{\cellcolor{gray!30}Case-II: Normal light neutrino ordering}\\
${\rm Hierarchies~\rightarrow }$&$M_{i+1}/M_i=10^2$&$M_{i+1}/M_i=10^3$ &$M_{i+1}/M_i=10^4$\\
\hline
${\rm Upper~bound~(GeV)}$&$2.57\times 10^{11}$&$2.58 \times 10^{11}$ &$2.59 \times 10^{11}$ \\
\hline
${\rm Lower~bound~~(GeV)}$&$2.50 \times 10^{11}$&$2.52\times 10^{11}$&$2.54\times 10^{11}$\\
\hline
 \multicolumn{4}{|c|}{\cellcolor{gray!30}Case-II: Inverted light neutrino ordering}\\
${\rm Hierarchies~\rightarrow }$&$M_{i+1}/M_i=10^2$&$M_{i+1}/M_i=10^3$ &$M_{i+1}/M_i=10^4$\\
\hline
${\rm Upper~bound~(GeV)}$&$5.38\times 10^{11}$&$5.40 \times 10^{11}$ &$5.42\times 10^{11}$ \\
\hline
${\rm Lower~bound~~(GeV)}$&$5.25\times 10^{11}$&$5.27 \times 10^{11}$&$5.28 \times 10^{11}$\\
\hline
\end{tabular} 
\end{center} 
\end{table}
One can see from Table \ref{t8} that the lower and upper bounds on $M_1$ slightly differ in each hierarchical cases. As explained before, the  first term in Eq.\eqref{epsim} is not sensitive to the chosen hierarchies. However, the second term contributes to the $\varepsilon_1^\mu$ and hence to the final $Y_B$. Thus for the same value of $M_1$, contribution from the second term in Eq.\eqref{epsim} is larger for $M_{i+1}/M_i=10^{2}$ and smaller for $M_{i+1}/M_i=10^{4}$ compare to $M_{i+1}/M_i=10^{3}$ case. Hence for  $M_{i+1}/M_i=10^{2}$ case, slope of the $Y_B$ Vs. $M_1$ curve is larger  than the  case of $M_{i+1}/M_i=10^{3}$. Consequently both  upper and the lower bounds get slightly lowered (compared to standard $M_{i+1}/M_i=10^{3}$ case) for the given range of $Y_B$. Proceeding in the same way we obtain a little bit increased bounds for $M_{i+1}/M_i=10^{4}$ case.\\

\noindent
\underline{ Direct effect of $N_{2}$:}\\  

\noindent
For simplicity, here we consider only the effect of $N_2$.  It is shown in Ref.\cite{n2lp} that, due to a decoherence effect, a finite amount of lepton asymmetry generated by $N_2$ decays  get protected against  $N_1$-washout thus survive down to the electroweak scale and  contribute to the final
\begin{figure}[H]
\begin{center}
\includegraphics[scale=0.4]{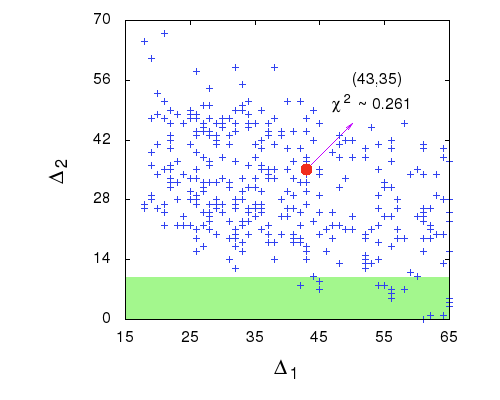}\includegraphics[scale=0.4]{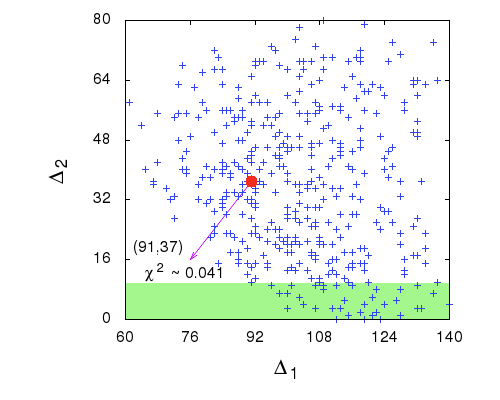}\\
\caption{Plots of the wash-out parameters $\Delta_1$ and $\Delta_2$ for inverted light neutrino mass ordering for both the cases. The red dot corresponds to the corresponding $\chi^2_{min}$ for which we calculte the final baryon aqsymmetry.}\label{n2}
\end{center}
\end{figure}
\noindent
 baryon asymmetry. For this procedure to happen, two wash-out parameters $\Delta_{1}=\frac{h_{11}}{M_1m^*}$ and $\Delta_2=\frac{h_{22}}{M_2m^*}$ must satisfy the condition $\Delta_1\gg 1\hspace{1mm} {\rm and}\hspace{1mm}\Delta_2 \not\gg 1$ with  $m^*=1.66\sqrt{g^*}\pi v^2/M_{Pl}\approx 10^{-3}$ eV. Here $\Delta_1\gg 1$ indicates that faster $N_1$ interactions destroy  coherence among the states produced by $N_2$, thus a part of the lepton asymmetry produced by $N_2$ survives orthogonal to $N_1$-states and gets protected against  $N_1$-washout. On the other hand, a mild wash-out of the lepton asymmetry produced by $N_2$ due to $N_2$-related interactions is represented by  $\Delta_2 \not\gg 1$ condition. For this mild wash-out scenario,  a sizable $N_2$-generated lepton asymmetry survives during the $N_1$-leptogenesis phase. It has been found that for each of the discussed cases, for a normal light neutrino mass ordering, both the wash-out parameters $\Delta_{1,2}<10$. Thus faster $N_1$ interaction do not take place and  condition for $N_2$ leptogenesis is violated. On the other hand for inverted light neutrino  mass orderings, the allowed parametric region  prefers large values of $\Delta_2$ in excess of 10 except at the bottom (green band). Thus the $\Delta_2 \not\gg 1$ condition is  violated in most of the region. Moreover the $\chi^2_{min}$ values, for which we calculate final $Y_B$ strongly violates $\Delta_2 \not\gg 1$ condition. Few allowed points with $\Delta_2<10$ correspond to values of $\chi^2$ above 0.8 which is much higher than $\chi^2_{min}$ for which we obtain $Y_B$ in the observed range. Therefore, for our calculation, any direct effect of $N_2$ is not significant. Note that 0.8 is not a special value. What we are trying to address, is that there are some data points in the model parameter space for which the conditions for $N_2$ leptogenesis is satisfied. However, the minimum value of $\chi^2$ for those data sets is 0.8. This means the corresponding observables are much away from their best-fit values. Practically  every data point in the parameter space is acceptable if they produce $Y_B$ in the observed range. However, throughout the analysis, for the computation of $Y_B$,  we restrict ourselves close to the best fitted values. In that sense the data points which are away from the best fit values are disfavored.

\section{Summary and conclusion}\label{s7}

We present the Strong Scaling Ansatz (SSA) as a residual $\mathbb{Z}_2\times\mathbb{Z}_2$ symmetry. Since SSA predicts a vanishing $\theta_{13}-$thus no Dirac CP violation, we  modify SSA with a complex extension of the residual $\mathbb{Z}_2\times\mathbb{Z}_2$ by invoking a nonstandard CP transformation and address the new symmetry as a generalized $\mathbb{Z}_2\times\mathbb{Z}_2$ symmetry. Depending upon the implementation of the symmetry, there are several cases that have been explored in this model.  For each of the cases, besides the  predictions of maximal Dirac CP violation ($\delta=\pm \pi/2$) and CP conserving values for the Majorana phases ($\alpha,\beta=0,\pi$), constrained ranges for the $\beta\beta0\nu$ decay parameter $|M_{ee}|$ and the light neutrino masses are also found. In this extended SSA, both the neutrino mass orderings are found to be allowed with upper bounds on $\sum_{i}m_i$ that are much lower than the present value $0.23$ eV.\\

We further discuss the  generalized $\mathbb{Z}_2\times\mathbb{Z}_2$ within the framework of type-I seesaw mechanism. Baryogenesis via leptogenesis scenario has been explored qualitatively as well as quantitatively. Typical  structure
of the Dirac mass matrix $m_D$ leads to a common origin of $\theta_{13}$, leptonic CP violation and nonzero CP asymmetry parameter  $\varepsilon_i^\alpha$. Here we focus the $N_1$-leptogenesis scenario as the primary one. However, we also  discuss the effect of the heavier neutrinos $N_{2,3}$ on the final baryon asymmetry $Y_B$. It is shown that the heavier neutrinos might effect the final $Y_B$ in two ways i) via the chosen hierarchy of the RH neutrinos and ii) through the asymmetry generated by the heavy neutrino itself (for simplicity we have assumed only the effect of $N_2$, i.e., $N_2$ leptogenesis). We found that the final $Y_B$ is not sensitive to  the chosen hierarchy of the RH neutrinos since the leading order term in $\varepsilon_1^\alpha$ is independent of the chosen hierarchy. Throughout the analysis we restrict ourselves to the near best-fit values of the oscillation parameters for which a positive value of $Y_B$ is obtained. We found that the conditions for $N_2$ leptogenesis are not satisfied for those best-fit points. Thus $N_2$-leptogenesis is also not so sensitive to the final $Y_B$. For each of the cases and irrespective of the light neutrino mass ordering, only $\tau$-flavored leptogenesis  scenario ($10^9$ GeV $<T\sim M_1 <10^{12}$ GeV) is found to be feasible one to generate $Y_B $ in the observed range with the other regimes $T\sim M_1>10^{12}$  and  $T\sim M_1<10^9$ GeV being ruled  out analytically as well as numerically. The best-fit parameters for which we calculate the final $Y_B$, lead to the value of $\theta_{23}<45^0$ for normal  mass orderings and $\theta_{23}>45^0$ for inverted mass orderings for both Case-I and Case-II. We also found an upper and a lower bound on the lightest ($M_1$) of the heavy neutrino masses for each case. Finally for a fixed value of $M_1$ we also investigate $\theta_{23}$ sensitivity of the final $Y_B$. Although both the light neutrino mass orderings are allowed, the normal mass ordering comes up with an interesting prediction. It has been shown and  explained in Sec.\ref{s5} that in both the normal mass ordering scenarios, there exist  lower limits on $M_1$, above which any value of $M_1$ corresponds to $\theta_{23}<45^0$ for $Y_B$ to be in the observed range. \\

As a final note, the predictions of this model--thus the viability of  modification to SSA with a generalized $\mathbb{Z}_2\times\mathbb{Z}_2$ symmetry,  would be tested  in the ongoing experiments such as GERDA-II\cite{gerda2}, T2K\cite{t2k}, NO$\nu$A\cite{noa} etc. shortly.
 \section*{Acknowledgment} The work of the authors are supported by the the Department of Atomic Energy (DAE), Government of India. The authors would like to thank Probir Roy for discussion about residual symmetry. R. Samanta would like to thank Walter Grimus for a valuable discussion on CP transformation during a visit at the University of Vienna. 
%\section{hi}

\end{document}